\documentclass{aa}  
\pdfoutput=1
\usepackage{graphicx}
\usepackage{txfonts}
\usepackage[colorlinks,citecolor=blue,urlcolor=blue,filecolor=blue,linkcolor=blue]{hyperref}
\usepackage{amsmath}
\usepackage{amssymb}
\usepackage{upgreek}
\usepackage{natbib}
\usepackage{xcolor}
\usepackage{enumitem}
\usepackage[normalem]{ulem}

\begin{document} 

\title{Searching for pulsars associated with polarised point sources using LOFAR: Initial discoveries from the TULIPP project  }
\titlerunning{TULIPP survey description and initial discoveries}
\authorrunning{C.\ Sobey et al.}

\author{C.~Sobey\inst{1}
  \and
  C.~G.~Bassa\inst{2}
  \and
  S.~P.~O'Sullivan\inst{3}
  \and
  J.~R.~Callingham\inst{4,2}
  \and
  C.~M.~Tan\inst{5,6}
  \and
  J.~W.~T.~Hessels\inst{2,7}
  \and
  V.~I.~Kondratiev\inst{2,8}
  \and 
  B.~W.~Stappers\inst{9}
  \and
  C.~Tiburzi\inst{2}
  \and 
  G.~Heald\inst{1}
  \and
  T.~Shimwell\inst{2,4}
  \and
  R.~P.~Breton\inst{9}
  \and
  M.~Kirwan\inst{3}
  \and
  H.~K.~Vedantham\inst{2,10}
  \and
  Ettore~Carretti\inst{11}
  \and
  J.~-M.~Grie{\ss}meier\inst{12,13}
  \and
  M.~Haverkorn\inst{14}
  \and
  A.~Karastergiou\inst{15}
  %\fnmsep\thanks{Just to show the usage of the elements in the author field}
}

\institute{CSIRO Space and Astronomy, PO Box 1130 Bentley, WA 6102, Australia\\
  \email{sobey.charlotte@gmail.com}
  \and
  ASTRON, Netherlands Institute for Radio Astronomy, Oude Hoogeveensedijk 4, 7991 PD, Dwingeloo, The Netherlands\\
  \email{bassa@astron.nl}
  \and
  School of Physical Sciences and Centre for Astrophysics \& Relativity, Dublin City University, \mbox{D09 W6Y4 Glasnevin, Ireland}
  \and
  Leiden Observatory, Leiden University, PO Box 9513, 2300 RA, Leiden, The Netherlands
  \and
  Department of Physics, McGill University, 3600 rue University, Montr\'{e}al, QC H3A 2T8, Canada
   \and
   McGill Space Institute, McGill University, 3550 rue University, Montr\'{e}al, QC H3A 2A7, Canada
   \and
   Anton Pannekoek Institute for Astronomy, University of Amsterdam, Science Park 904, 1098 XH, Amsterdam, The Netherlands
   \and
   Astro Space Centre, Lebedev Physical Institute, Russian Academy of Sciences, Profsoyuznaya Str. 84/32, Moscow 117997, Russia
   \and
   Jodrell Bank Centre for Astrophysics, School of Physics and Astronomy, University of Manchester, Manchester, M13 9PL, UK
   \and
   Kapteyn Astronomical Institute, University of Groningen, Postbus 800, 9700 AV Groningen, The Netherlands
   \and
   INAF Istituto di Radioastronomia, Via Gobetti 101, 40129 Bologna, Italy
   \and
   LPC2E - Universit\'{e} d'Orl\'{e}ans / CNRS  
   \and
   Station de Radioastronomie de Nan\c{c}ay, Observatoire de Paris, CNRS/INSU, USR 704 - Univ. Orléans, OSUC, 18330 Nan\c{c}ay, France
   \and
   Department of Astrophysics/IMAPP, Radboud University, PO Box 9010, 6500 GL, The Netherlands
   \and
   Oxford Astrophysics, Denys Wilkinson Building, Keble Road, Oxford OX1 3RH, UK
  %\thanks{The university of heaven temporarily does not accept e-mails}
}

\date{Received December 11, 2021; accepted February 9, 2022}

\abstract{
Discovering radio pulsars, particularly millisecond pulsars (MSPs), is important for a range of astrophysical applications, such as testing theories of gravity or probing the magneto-ionic interstellar medium. We aim to discover pulsars that may have been missed in previous pulsar searches by leveraging known pulsar observables (primarily polarisation) in the sensitive, low-frequency radio images from the Low-Frequency Array (LOFAR) Two-metre Sky Survey (LoTSS), and have commenced the Targeted search, using LoTSS images, for polarised pulsars (TULIPP) survey. For this survey, we identified linearly and circularly polarised point sources with flux densities brighter than 2\,mJy in LoTSS images at a centre frequency of 144\,MHz with a 48\,MHz bandwidth. Over 40 known pulsars, half of which are MSPs, were detected as polarised sources in the LoTSS images and excluded from the survey. We have obtained beam-formed LOFAR observations of 30 candidates, which were searched for pulsations using coherent de-dispersion. Here, we present the results of the first year of the TULIPP survey. We discovered two pulsars, PSRs\,J1049+5822 and J1602+3901, with rotational periods of $P=0.73$\,s and 3.7\,ms, respectively. We also detected a further five known pulsars (two slowly-rotating pulsars and three MSPs) for which accurate sky positions were not available to allow a unique cross-match with LoTSS sources. This targeted survey presents a relatively efficient method by which pulsars, particularly MSPs, may be discovered using the flexible observing modes of sensitive radio telescopes such as the Square Kilometre Array and its pathfinders/precursors, particularly since wide-area all-sky surveys using coherent de-dispersion are currently computationally infeasible.
%\LEt{Please avoid the use of the slashes. For details, refer to Sect. 2.9 of the language guide.
%Please check for this throughout.}
}


\maketitle

\section{Introduction}\label{sec:introduction}

Discovering pulsars is vital for improving our understanding of the Galactic population of neutron stars \citep[e.g.][]{Lorimer+2006,Kaspi+2018,Rajwade+2018}.
In particular, the discovery of millisecond pulsars (MSPs; with rotational spin periods of $P\lesssim10$\,ms, also referred to as `recycled' pulsars; e.g.\ \citealt{Tauris+2012}) is a high priority for pulsar surveys because they provide particularly unique and powerful probes of fundamental physics \citep[e.g.][]{Demorest+2010,Antoniadis+2013,Ransom+2014,Archibald+2018}, including towards detecting nanohertz gravitational wave signals via Pulsar Timing Arrays \citep[e.g.][]{Perera+2019}.
In addition, a dense network of pulsars provides three-dimensional lines of sight, which allows us to investigate our Galaxy's intervening magneto-ionic interstellar medium
\citep[ISM; e.g.][]{ymw17,Deller+2019,sbg+19}. 

%discovering pulsars
To date, the majority of the pulsar population, and most slowly rotating (non-recycled) radio pulsars, have been discovered using wide-area, untargeted surveys, where large areas of the sky are observed at high time resolution to search for periodic, dispersed signals. These surveys have used either large single dish antennas \citep[e.g.][]{PMPS,PALFA,HTRU,dsm+13,GBNCC,SUPERB,Cameron+2020,Han+2021},
or interferometers \citep[e.g.][]{GHRSS,MeerTRAP,VenkatramanKrishnan+2020,Good+2021,Swainston+2021}.
Moreover, a large number of MSPs have been discovered through targeted surveys towards globular clusters\footnote{\url{http://www.naic.edu/~pfreire/GCpsr.html}} \citep[e.g.][]{Ransom+2005,Hessels+2006,Hessels+2007,Ransom_2008,Dai+2020,Pan+2021,Ridolfi+2021} and/or high-energy sources 
\citep[e.g.][]{Wijnands+1998,Abdo+2010,Camilo+2015,bph+17,Clark+2017,Frail+2018}.
%discovery - images
Several pulsars, including the first MSP known, have been discovered through identification as pulsar candidates using radio interferometric images, selected based on criteria including a high degree of polarisation, steep spectrum, and/or variability \citep[e.g.][]{Backer+1982,Hamilton+1985,Lyne+1987,Strom1987,Kulkarni+1988,Navarro+1995,Jelic+2015,Kaplan+2019}. 

%coherent dedispersion
The impulsive radio signals from pulsars are delayed in time as they propagate through the ionised ISM, proportionally to the observing wavelength squared, an effect known as dispersion.
To accurately detect the pulse profiles of pulsars, this effect must be corrected in radio observation data through dispersion removal, or de-dispersion \citep[e.g.][]{Lorimer+Kramer}.
The data from all-sky pulsar surveys require de-dispersion at many trial dispersion measures (DMs) up to thousands of pc\,cm$^{-3}$ \citep[e.g.][]{Lyon+2016}.
Typically, incoherent de-dispersion has been used, which is a computationally expensive process, although recent implementations for graphics processing
units (GPUs) have enabled real-time processing \citep[e.g.][]{Magro+2011,bbbf12,Sclocco+2020}.

Coherent de-dispersion can almost completely mitigate the effects of dispersive smearing of the pulse profile up to high dispersion measures \citep[e.g.][]{hr75} and is particularly important for low-frequency observations, due to the dependence on wavelength.
However, coherent de-dispersion requires larger data volumes, increasing the computational requirements, rendering its use in wide-area pulsar surveys computationally infeasible at present \citep[e.g.][]{bph17}. 
However, its use in targeted searches of pulsar candidate sources is feasible.
For example, a targeted survey using Low-Frequency Array (LOFAR; \citealt{sha+11,hwg+13}) observations of unidentified $\gamma$-ray sources from the \textit{Fermi} $\gamma$-ray Space Telescope \citep{fermi} pioneered the use of GPU-accelerated coherent dispersion removal for pulsar searches at low frequencies \citep{bph17}.
This resulted in the discovery of three fast-spinning ($P<5$\,ms) MSPs \citep[e.g.][]{Pleunis+2017}, including the fastest-spinning MSP known in the Galactic field (i.e. outside a globular cluster; $P=1.41$\,ms; \citealt{bph+17}). 
Two of these pulsars also show particularly steep radio spectral indices\footnote{We use the convention $S_{\nu}\propto\nu^{\alpha}$, where flux density, $S$, is proportional to observing frequency, $\nu$.} ($\alpha<-2.5$), and they likely could have only been discovered at low radio frequencies ($\nu<300$\,MHz).

%General pulsar properties - polarisation and spectrum
The emission from pulsars is often highly polarised (average linear polarisation fractions of $\sim$20\% that usually increase towards lower frequencies, and average circular polarisation fractions of $\sim$10\%\ that usually decrease towards lower frequencies), which provides valuable information about the intrinsic radio emission and its propagation through the ISM
\citep[e.g.][]{rc69,gl98,Johnston+2008,Noutsos+2015,Xue+2019,Sobey+2021}. 
In addition, the radio spectral indices of pulsar emission are generally described well by a steep power law \citep[e.g.][]{Bates+2013,Bilous+2016,Jankowski+2018}.
For example, 15 of 48 MSPs detected using LOFAR have spectral indices of $\alpha<-2$ \citep{Kondratiev+2016}.
For comparison, the average spectral indices for typical radio sources in large surveys is $\sim0.7-0.8$ \citep[e.g.][]{deGasperin+2018}.
Indeed, understanding the population of steep-spectrum, polarised, short-period pulsars is especially critical for determining the prospects and strategies for imaging, wide-area and targeted pulsar surveys using the Square Kilometre Array (SKA), which has the potential to discover thousands of pulsars and MSPs \citep[e.g.][]{Hessels+2015,Keane+2015,Xue+2017,Levin+2018}.

% Using low frequency LOFAR images
The LOFAR Two-metre Sky Survey (LoTSS\footnote{\url{https://lofar-surveys.org/surveys.html}}) is a deep 120--168\,MHz imaging survey that facilitates a range of science goals and will eventually cover the entire northern sky \citep{srb+17}. 
The first LoTSS public data release (DR1) covers 424\,deg$^{2}$ in the region of the HETDEX Spring Field with up to 6\arcsec\ resolution and 71\,$\upmu$Jy/beam median sensitivity \citep{sth+19}. 
The second LoTSS data release (DR2) covers a larger sky area of 5635\,deg$^{2}$ and includes full polarisation data products (Stokes $I,Q,U,V$) \citep{Tasse+2021,Shimwell+2022}.
The LOFAR Magnetism Key Science project\footnote{\url{https://lofar-mksp.org}} used the LoTSS-DR2 data products to produce an `\textsc{RM Grid}': a catalogue of $\sim$2,500 linearly polarised sources and their Faraday rotation measures (RMs; O'Sullivan et al., in prep.). 
This facilitates a range of science goals aimed at investigating magnetic fields
\citep[e.g.][]{Heald+2020,O'Sullivan+2020,HerreraRuiz+2021,Hutschenreuter+2022}.
Despite the high Galactic latitudes of the observations ($b\sim-33\degr$ \& $70\degr$), this catalogue contains over 30 known pulsars, over 50\% of which are MSPs (O'Sullivan et al., in prep.).
Furthermore, the circularly polarised component of LoTSS-DR2 (V-LoTSS; Callingham et al., in prep.) represents the most sensitive, wide-field Stokes $V$ survey conducted to date. 
There are tens of known pulsars and many highly circularly polarised sources with no optical counterparts.
LoTSS is the first imaging survey with sufficient sensitivity to detect most nearby radio pulsars as point sources. 
The linear and circular polarisation catalogues from the LoTSS survey, therefore, present a novel, sensitive, and efficient method to identify sources that may be undiscovered pulsars (candidates).
For confirmation, these sources can be followed-up at the same frequencies using LOFAR pulsar observations and searched for periodic signals. 

Particularly relevant to this work is the LOFAR Tied-Array All-Sky Survey (LOTAAS\footnote{\url{http://www.astron.nl/lotaas}}). 
This all-sky low-frequency (135\,MHz) pulsar survey, using the addition of the signals from the six central `Superterp' LOFAR High-Band Antenna (HBA) stations, has discovered 74 pulsars to date \citep{Coenen+2014,tbc+18,Sanidas+2019,Michilli+2020,Tan+2020}.
The time sampling (492\,$\upmu$s) and incoherent de-dispersion used in processing the LOTAAS data limits the sensitivity to MSPs with $\mathrm{DM}>10$\,pc\,cm$^{-3}$ \citep{Sanidas+2019}. 
For comparison, 96\% of the known MSP population have $\mathrm{DMs}>10$\,pc\,cm$^{-3}$, with a median of $57$\,pc\,cm$^{-3}$ \citep[ATNF Pulsar Catalogue;][]{mhth05}\footnote{psrcat version 1.65 accessed at \url{ http://www.atnf.csiro.au/research/pulsar/psrcat/}}.

%complementarity - LOFAR
Targeted searches complement and enhance the capabilities of LOFAR pulsar surveys to discover pulsars. 
Not only can these make use of coherent de-dispersion, but greater observing sensitivities may also be achieved.
Depending on the positional uncertainties of candidate sources, additional LOFAR HBA stations can be included in the tied-array observations, up to the innermost 24 core stations (longest baseline 3.6\,km), providing $\approx$4$\times$ greater instantaneous sensitivity than six `superterp' stations (longest baseline 0.3\,km) alone.
Targeted pulsar searches (including selecting promising candidates using LoTSS) are also beneficial because the source's sky positions are known to a higher precision ($\approx0\farcs2$; \citealt{Shimwell+2022}) compared to those obtained from traditional, wide-area pulsar surveys, which aids achieving a pulsar timing solution more efficiently (e.g. sky position is covariant with rotational spin-down, $\dot{P}$; e.g. \citealt{Parthasarathy+2019}). 

%Thus, the TULIPP  survey.
Here, we present the methodology and the results from the first year (two LOFAR observing cycles, LC12 and LC13, 2019 June 1 -- 2020 May 31) of the ongoing targeted radio pulsar survey, TULIPP (Targeted search, Using LoTSS Images, for Polarised Pulsars).
%\LEt{Please remember to spell out acronyms upon first appearance, \uline{\textbf{followed by the
%acronym in parentheses}}, in the abstract and then again beginning with the introduction. After this,
%only use the acronym. Please check carefully throughout.}.
In \S\,\ref{sec:selection} we describe the selection of promising pulsar candidates, identified as linearly and circularly polarised LoTSS unresolved/point sources.
%\LEt{reminder note 2}. 
In \S\,\ref{sec:observations}, we provide an overview of the LOFAR pulsar observations using the high time resolution, tied-array, beam-forming modes and the data processing including the pulsar search pipelines. 
We present our results in \S\,\ref{sec:results}, including the first two pulsars discovered in the TULIPP survey, PSRs\,J1049+5822 and J1602+3901.
We discuss and conclude in \S\,\ref{sec:discussion} and \S\,\ref{sec:conclusions}.

The TULIPP survey is now a long-term LOFAR project, and the number of pulsar candidates has continued to grow after processing additional LoTSS DR2 fields. The candidates and results from LC14 onward will be presented in a follow-up paper (Bassa et al. in prep.).

\section{Selection of pulsar candidates}\label{sec:selection}

%LoTSS fields processing and initial candidate selection
The pulsar candidate selection matured in line with the progress of the LoTSS data processing for the linear and circular polarisation images. 
Prior to LC12 and LC13, $\sim150$ and $\sim845$ LoTSS fields had been processed through the linear polarisation \textsc{`RM Grid'} pipeline and catalogued in circular polarisation, respectively.  
These produced a catalogue of over $\sim2500$ and $\sim100$ sources, respectively (O'Sullivan et al., in prep.; Callingham et al., in prep.). 
During LC12 and LC13, we observed nine linearly polarised pulsar candidates, and 21 circularly polarised pulsar candidates, summarised in Table~\ref{tab:sources}. 
We describe how these sources were selected below.

% Selection criteria para - linear
We used the following initial criteria to identify pulsar candidates in linear polarisation images from the LoTSS \textsc{RM Grid} catalogue:
%\LEt{A\&A does not allow itemised listing outside of your conclusions. Please put these points into the paragraph}
%\begin{enumerate}[label=(\roman*)]
1) unresolved in LoTSS linear polarisation images at $20\arcsec$ resolution (later confirmed using the Stokes $I$\ LoTSS images with $6\arcsec$ resolution). Since pulsars are compact objects of only $\sim$tens of kilometres in size \citep[e.g.][]{Riley+2021} at $\sim$kpc distances, they appear as point sources in images. 
%\LEt{please only use the symbol with figures}
2)  $>$2\,mJy total intensity (Stokes $I$) flux density. This limit provides LoTSS sources detected with sufficient sensitivity ($\sim30\upsigma$) to obtain reliable polarisation measurements, and they are suitably bright to be detected using LOFAR Core-station, tied-array, beam-formed observations of $\sim20-80$\,mins. For comparison to higher frequency surveys, this limit corresponds to 20\,$\upmu$Jy at 1.4\,GHz for a source with a spectral index of $-2$.
3) $>$7\%\ linear polarisation fraction. This is both greater than the expected instrumental polarisation leakage in LoTSS images \citep{Shimwell+2022}, and higher than many radio galaxies and blazars at 144\,MHz, which have larger |RMs| and de-polarisation due to Faraday complexity \citep[e.g.][O'Sullivan et al., in prep.]{Anderson+2015,Lenc+2017}.
%\end{enumerate}

% Selection criteria para - circular
We used the following initial criteria to identify pulsar candidates in circular polarisation data from the V-LoTSS catalogue: 
%\begin{enumerate}[label=(\roman*)]
1) unresolved in the Stokes $I$\ LoTSS images with 6\arcsec\ resolution.
2) $>$5\,mJy total intensity flux density. There were a larger number of candidate sources in circular polarisation than linear polarisation, so we used a higher flux density threshold than that for the linearly polarised sources to obtain a sufficient number of candidates to begin the TULIPP survey. 
3) $>$1\%\  absolute (i.e. left or right handed) circular polarisation fraction. This is greater than the expected instrumental polarisation leakage \citep{Callingham+2021} and also higher than many radio galaxies and blazars \citep[e.g.][]{Macquart+2004}.
%\end{enumerate}

% Selection criteria para - cross matching sources - PSRcat
These initial candidate lists were cross-checked with the ATNF Pulsar Catalogue, as well as pulsar discovery websites for ongoing surveys introduced in \S\,\ref{sec:introduction}.
Any source positions cross-matched with those of known pulsars were removed from the candidate list.
This resulted in the removal of 25 linearly polarised sources in total for LC12 and LC13 (13 MSPs; 3 mildly recycled pulsars; 8 non-recycled pulsars).
Nine of these pulsars have RMs published in the literature and show good agreement with the LoTSS \textsc{RM Grid} results \citep[e.g.][O'Sullivan et al., in prep.]{mhth05,sbg+19}.
This also resulted in the removal of 35 circularly polarised sources coincident with known pulsars (approximately half of which are MSPs).

% Selection criteria para - cross matching sources - multi-wavelength
The remaining pulsar candidates were cross-matched for multi-wavelength counterparts within a conservative search radius of 10\arcsec\ from the LoTSS source positions in the optical (including SDSS; \citealt{aaa+15sdss}, GAIA; \citealt{bvp+16,bvp+18}, and Pan-STARRS; \citealt{cmm+16}), infrared (including 2MASS; \citealt{csd+03}, SWIRE; \citealt{Lonsdale+2003}, WISE; \citealt{Wright+2010}), radio (including NVSS; \citealt{ccg+98}, FIRST; \citealt{hwb+15}), and $\gamma$-ray (\textit{Fermi}; \citealt{fermi+19}).
We refined the list of candidates such that they either have no associated multi-wavelength counterpart within $10\arcsec$ in these surveys, or their multi-wavelength counterparts do not preclude the possibility that they are pulsars or pulsar systems, including optical/IR counterparts fainter than $r=16$ and not classified as a galaxy. This reduced the list of candidates by more than half, while preventing selecting against pulsars with optically bright binary companions \citep[e.g.][]{Antoniadis+2021}.
NVSS and FIRST surveys were used to exclude sources because these are likely to be radio galaxies with flatter spectral indices.
No known pulsars were excluded using the NVSS or FIRST surveys' cross-match \citep[e.g.][]{Kaplan+1998}, as these had already been deselected using the ATNF pulsar catalogue.

\begin{figure*}
        \includegraphics[width=\columnwidth]{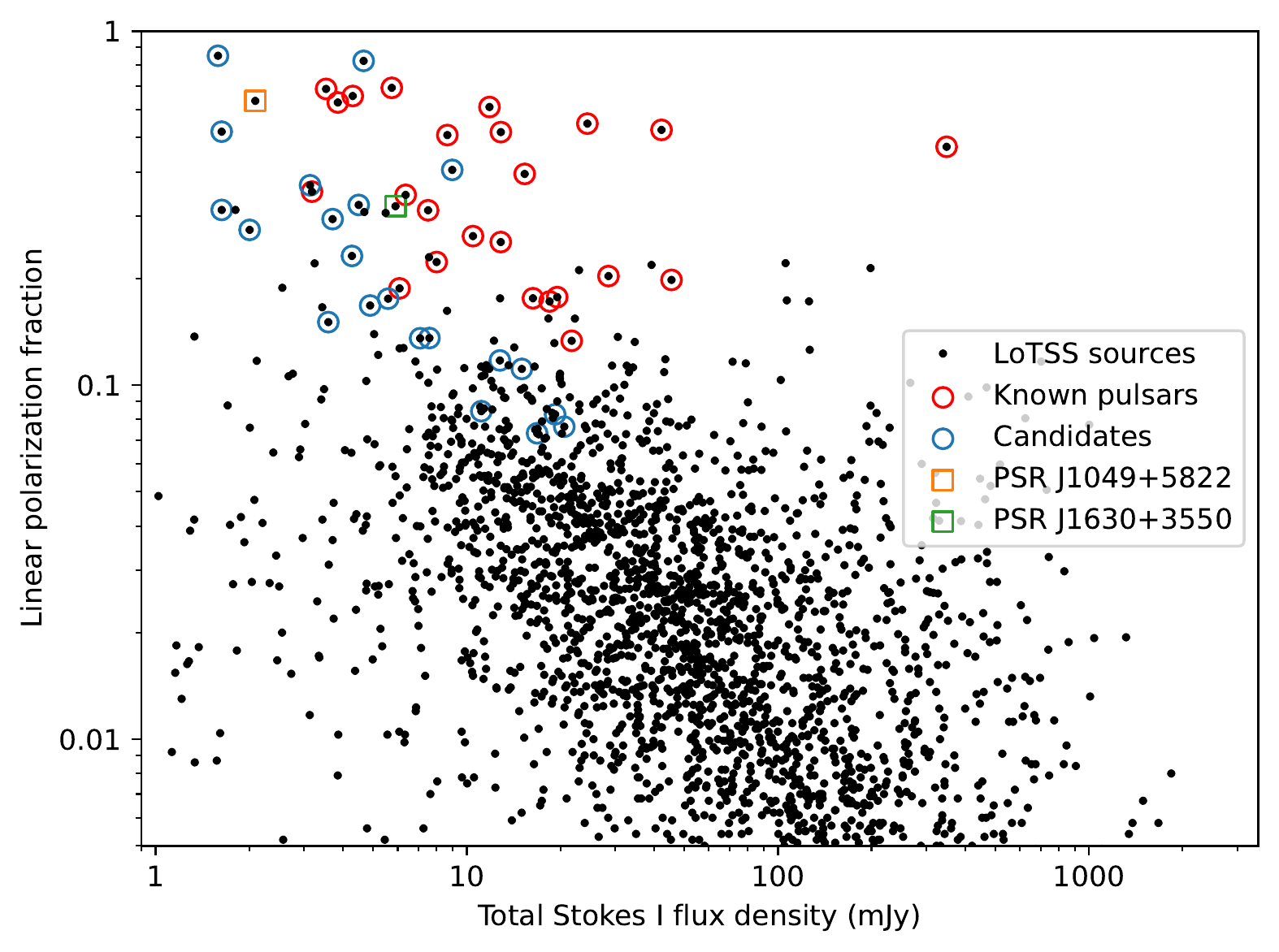}
        \includegraphics[width=\columnwidth]{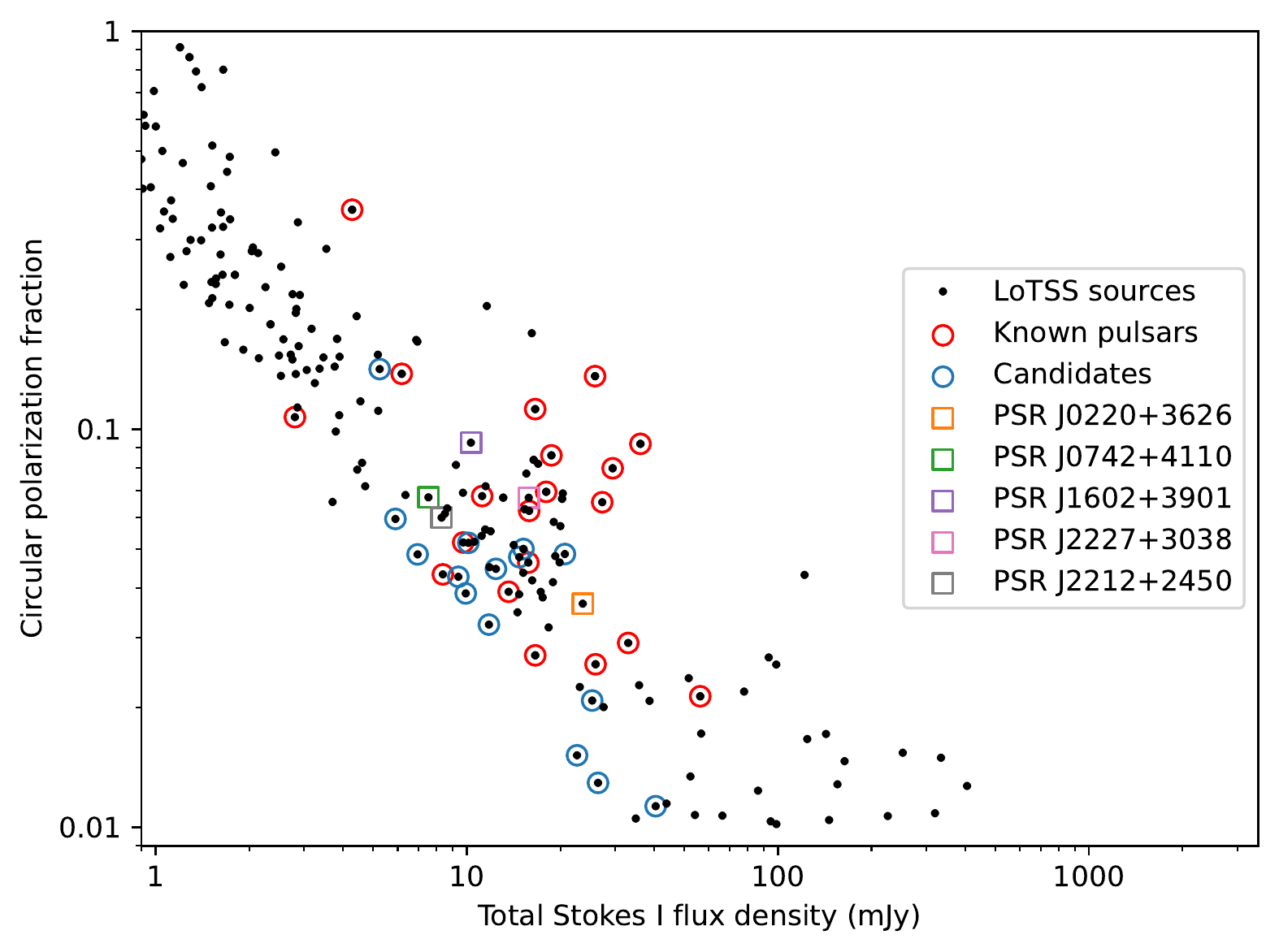}
    \caption{Fractional polarisation against total intensity (Stokes $I$) flux density for: LoTSS sources (black points); pulsars discovered or detected through the TULIPP survey (squares);  candidates observed and searched for pulsars in this work (blue circles); and known pulsars (red circles). Left: linearly polarised sources from the \textsc{RM Grid} (O'Sullivan et al., in prep.). Right: circularly polarised sources (absolute values) from V-LoTSS (Callingham et al., in prep.). }
    \label{fig:Lsummary}
\end{figure*}

%Final candidates and table
The properties of the nine selected linearly polarised sources include Stokes $I$ flux densities at 144\,MHz in the range of $S_{144}=2$ to 20\,mJy; linear polarisation fractions from 7 to 82\%; and absolute RMs in the $2<|\rm{RM}|<73$\,rad\,m$^{-2}$ range, Table~\ref{tab:sources}.
The LoTSS frequency range ($120-168$\,MHz) and channel bandwidth (97.6\,kHz) provide a nominal resolution in Faraday depth space (i.e. FWHM of the rotation measure spread function, RMSF; \citealt{Brentjens+2005}) of 1.15\,rad\,m$^{-2}$.
The maximum observable $\mathrm{|RM|=170}$\,rad\,m$^{-2}$ (at full sensitivity), although the Faraday depth range in the \textsc{RM Grid} RM synthesis processing was limited to $\mathrm{|RM|=120}$\,rad\,m$^{-2}$ (O'Sullivan et al., in prep.).
The fractional linear polarisations were computed as the linearly polarised flux density ($L=\sqrt{Q^{2}+U^{2}}$, where $Q$ and $U$ represent the Stokes $Q$ and Stokes $U$ flux densities, after Faraday rotation correction) divided by the total intensity flux density (Stokes $I$) obtained from the total 8-h integration time. 

The circularly polarised sources have Stokes $I$ flux densities at 144\,MHz in the range of $S_{144}=5$ to 41\,mJy and circular polarisation fractions from 2\%\ to 22\%, Table~\ref{tab:sources}.
The fractional circular polarisations were computed as the circularly polarised flux density (Stokes $V$) divided by the total intensity flux density (Stokes $I$) obtained from the total 8-h integration time. 

We estimated the maximum DMs expected for the candidate pulsars using Galactic electron density distribution models NE2001 \citep{cl02} and YMW16 \citep{ymw17}; see Table~\ref{tab:sources}). 
These are usually below 50\,pc\,cm$^{-3}$ for the candidates at high Galactic latitudes of $|b|\gtrsim30\degr$. 
The largest DM predicted, 333\,pc\,cm$^{-3}$, is towards ILT\,J003904.45+625711.4 and close to the Galactic plane, $b=0.1\degr$.
We also estimated the largest RM expected due to Galactic Faraday rotation using the reconstruction\footnote{\url{https://wwwmpa.mpa-garching.mpg.de/~ensslin/research/data/faraday2020.html}} from \citet{Hutschenreuter+2022}; see Table~\ref{tab:sources}).
The largest |RM| predicted, $-204$\,rad\,m$^{-2}$, is also towards ILT\,J003904.45+625711.4.
In addition, we note that the LoTSS sources ILT\,J003904.45+625711.4 and ILT\,J024831.01+423020.3 are coincident with the positions of $\gamma$-ray sources 4FGL\,J0039.1+6257 and 4FGL\,J0248.6+4230, respectively, \citep{fermi+19} and these are further discussed in Section \ref{subsec:nondet}.

Figure~\ref{fig:Lsummary} shows the fractional linear (left panel) or circular (right panel) polarisation plotted against total intensity flux density for all of the polarised LoTSS sources from the \textsc{RM Grid} (O'Sullivan et al., in prep.) or V-LoTSS (Callingham et al., in prep.). We highlight the pulsars discovered and reported in this work, along with known pulsars from the pulsar catalogue or detected in this work, and candidate sources from Table~\ref{tab:sources}. This highlights the pulsar population that occupies the high fractional polarisation and relatively low flux density region of this parameter space. PSR\,J0218+4232 is the notable outlier in the linearly polarised sources, with a total intensity of $S_{144}=348.7$\,mJy and polarisation fraction of 47\%. This is an MSP with $P=2.3$\,ms and relatively high DM, 61.2\,pc\,cm$^{-3}$, for this sample of pulsars, with a steep spectral index \citep[e.g.][]{Kondratiev+2016}. This source was also first identified as a steep-spectrum, highly linearly polarised point source in interferometric images \citep{Erickson_1980,Navarro+1995}.

\begin{table*}
  \centering
  \footnotesize
  \caption{LoTSS sources selected as pulsar candidates based on their polarisation properties.}
  \label{tab:sources}
  \begin{tabular}{llrrrrrrrr}
    \hline\hline
    LoTSS source & Field & \multicolumn{1}{c}{$l$} & \multicolumn{1}{c}{$b$} & \multicolumn{1}{c}{$S_{144}$} & \multicolumn{1}{c}{p} & \multicolumn{1}{c}{$\mathrm{RM}$} & \multicolumn{1}{c}{$\mathrm{DM}_\mathrm{max}^\mathrm{NE2001}$} & \multicolumn{1}{c}{$\mathrm{DM}_\mathrm{max}^\mathrm{YMW16}$} &
    \multicolumn{1}{c}{$\mathrm{RM}_\mathrm{max}^\mathrm{H2020}$}\\
     & & \multicolumn{1}{c}{(\degr)} & \multicolumn{1}{c}{(\degr)} & \multicolumn{1}{c}{(mJy)} & \multicolumn{1}{c}{(\%)} & \multicolumn{1}{c}{(rad\,m$^{-2}$)} & \multicolumn{2}{c}{(pc\,cm$^{-3}$)} & \multicolumn{1}{c}{(rad\,m$^{-2}$)}\\
    \hline
    Linearly polarised  &  & &  &  & $L/I$ &  & & \\
    \hline
    \object{ILT\,J024831.01$+$423020.3}*            & P044+44 & $144.88$ & $-15.32$ & $  9.0$ & $41  $ & $  -37.8 $ & $94.8$ & $103.9$ & $-52$ \\
    
    \object{ILT\,J042737.00$+$684536.9}            & P068+69 & $140.49$ & $+13.63$ & $ 16.0$ & $7  $ & $  -72.7 $ & $103.6$ & $137.4$ & $-44$ \\
    
    \object{ILT\,J095618.49+461712.5}           & P149+47 & $171.92$ & $+50.68$ & $  5.7 $ & $82  $ & $  8.1 $ & $37.2$ & $24.9$  & $5$\\
    
    \object{ILT\,J104140.11+542419.3}            & P160+55 & $154.93$ & $+53.90$ & $  1.8 $ & $31  $ & $  2.1 $ & $34.6$ & $24.0$ & $2$ \\
    
    \object{ILT\,J104937.86+582217.6}$\star$           & P161+60 & $148.65$ & $+52.24$ & $  2.0 $ & $64  $ & $  4.0 $ & $35.0$ & $25.0$ & $10$ \\
    
    \object{ILT\,J105132.39+591355.8}            & P161+60 & $147.36$ & $+51.83$ & $  5.6 $ & $17  $ & $  8.2 $ & $35.1$ & $25.2$ & $7$ \\
    
    \object{ILT\,J140326.96+490750.2}            & P211+50 & $ 95.48$ & $+63.98$ & $ 23.6 $ & $8  $ & $  12.7 $ & $30.0$ & $22.2$ & $13$ \\
    
    \object{ILT\,J163035.93+355042.5}$\dag$  & P249+38 & $ 57.88$ & $+43.08$ & $   5.9 $ & $33  $ & $8.5 $ & $34.9$ & $29.5$ & $13$ \\
    
    \object{ILT\,J172307.05$+$651557.0}            & P260+65 & $ 95.17$ & $+33.65$ & $ 19.2$ & $8  $ & $  31.7 $ & $43.9$ & $38.0$  & $25$ \\
    
    \hline
    Circularly polarised  &  &  &  &  & $|V|/I$ &  &  & \\
    \hline
\object{ILT\,J003904.45$+$625711.4}  & P007+64 & $121.53$ & $ +0.12$ & $ 20.7$ & $ 5 \pm 2$ &  -- & $238.5$ & $332.7$  & $-204$ \\

\object{ILT\,J012902.97+191907.6}  & P024+21 & $135.04$ & $-42.68$ & $  62 $ & $ 22 \pm 6$ & -- & $39.8$ & $30.6$  & $-43$\\

\object{ILT\,J013942.10+312331.5}  & P024+31 & $134.87$ & $-30.36$ & $ 24.1 $ & $  2 \pm 1$ & -- & $51.6$ & $43.7$  & $-65$ \\

\object{ILT\,J022042.12+362655.7}$\dag$ &  P035+36 & $142.32$ & $-23.04$ & $ 24.5 $ & $ 4 \pm 1$ & -- & $66.4$ & $63.7$  & $-75$ \\

\object{ILT\,J024340.68+265149.7}  & P039+26 & $151.80$ & $-29.64$ & $ 15.4 $ & $  5 \pm 1$ & -- & $54.8$ & $49.4$  &  $-77$ \\

\object{ILT\,J074212.22$+$411015.0}$\dag$  & P114+39 & $178.13$ & $+26.57$ & $  7.5$ & $  7 \pm 2$ & -- & $63.1$ & $68.6$  & $6$ \\

\object{ILT\,J101650.83+433122.4}  & P153+42 & $174.86$ & $+54.89$ & $ 19.0 $ & $  3 \pm 1$ & -- & $35.3$ & $22.6$ & $5$ \\

\object{ILT\,J114600.31+504733.8}  & P16Hetdex13 & $146.25$ & $+63.27$ & $  8.6 $ & $  5 \pm 1$ & -- & $29.7$ & $21.1$  & $11$ \\

\object{ILT\,J115553.76+574020.9}  & P176+60 & $136.92$ & $+57.92$ & $ 236 $ & $  2 \pm 1 $ & -- & $32.3$ & $23.2$  & $5$ \\

\object{ILT\,J120106.46+284839.3}  & P181+27 & $201.40$ & $+78.77$ & $  18.5 $ & $  4 \pm 1$ & -- & $20.0$ & $18.9$  & $9$ \\

\object{ILT\,J120223.78+611016.4}  & P180+62 & $133.21$ & $+54.97$ & $  7.8 $ & $  6 \pm 1$ & -- & $33.2$ & $24.4$  & $14$ \\

\object{ILT\,J152745.28$+$545324.7}  & P232+55 & $ 88.28$ & $+50.39$ & $  9.4 $ & $  4 \pm 1$ & -- & $33.8$ & $26.2$  & $-2$ \\

\object{ILT\,J153514.64+483659.6}  & P231+50 & $ 78.38$ & $+51.86$ & $ 84.5 $ & $  2 \pm 1$ & -- & $33.5$ & $25.6$  & $7$ \\

\object{ILT\,J154721.26+493621.5}  & P236+48 & $ 78.90$ & $+49.66$ & $ 6.5$ & $  5 \pm 1$ & -- & $34.3$ & $26.4$  & $11$ \\

\object{ILT\,J160218.82+390101.5}$\star$  & P242+38 & $ 62.18$ & $+48.79$ & $ 27 $ & $  9 \pm 2$ & -- & $29.9$ & $26.6$  & $14$ \\

\object{ILT\,J163607.29+314631.5}  & P250+33 & $ 52.63$ & $+41.40$ & $ 16.3 $ & $  4 \pm 1$ & -- & $36.1$ & $30.6$  & $20$ \\

\object{ILT\,J164033.62+541943.4}  & P250+55 & $ 82.67$ & $+40.61$ & $ 31.0 $ & $  2 \pm 1$ & -- & $38.5$ & $31.2$  & $9$ \\

\object{ILT\,J181551.08+390739.3}  & P272+40 & $ 66.41$ & $+23.24$ & $ 33.7 $ & $  5 \pm 1$ & -- & $61.3$ & $51.5$  & $78$ \\

\object{ILT\,J212311.55$+$210431.7}  & P320+23 & $ 71.25$ & $-20.28$ & $  5.2$ & $  14 \pm 4 $ & -- & $68.3$ & $57.1$  & $-71$ \\

\object{ILT\,J221227.63+245036.7}$\dag$  & P333+26 & $ 82.96$ & $-25.43$ & $  8.3 $ & $  6 \pm 1$ & -- & $55.0$ & $46.9$  & $-82$ \\

\object{ILT\,J222741.77+303820.8}$\dag$  & P336+31 & $ 89.72$ & $-22.81$ & $  17.0 $ & $  7 \pm 2$ & -- & $60.5$ & $52.0$  & $-156$ \\

    \hline
  \end{tabular}
  \tablefoot{Columns provide the LoTSS catalogue source name, based on the right ascension and declination in the ILT\,Jhhmmss.ss$+$ddmmss.s format and with uncertainties $\approx0\farcs2$ \citep{Shimwell+2022}; LoTSS observing field identifier; source location in Galactic longitude and latitude; total Stokes $I$ flux density at 144\,MHz with 3\% median fractional uncertainty \citep{Shimwell+2022}; linear or circular fractional polarisation percentage (O'Sullivan et al., in prep.; Callingham et al., in prep ); RM measured from the RM Grid processing (only applicable to linearly polarised sources) with uncertainties $\approx0.1$\,rad\,m$^{-2}$ including ionospheric RM correction (O'Sullivan et al., in prep.); maximum DM expected for the source line of sight using Galactic electron density models: NE2001 and YMW16 (see \S\ref{sec:selection}); maximum Galactic RM expected for the source line of sight using the Faraday Sky 2020 reconstruction with 22\% median fractional uncertainty (see \S\ref{sec:selection}). LoTSS sources associated with the TULIPP pulsar discoveries or re-detections of known pulsars are noted using $\star$ and $\dag$, respectively, and they are also summarised in Table \ref{tab:pulsars}. Furthermore, the LoTSS source noted using * is coincident with PSR\,J0248+4230, recently discovered using the GMRT at 322\,MHz \citep{Bhattacharyya+2021} and further discussed in Section \ref{subsec:nondet}.   }
\end{table*}

\section{Observations and data processing}\label{sec:observations}

\subsection{Pulsar search observations}\label{subsec:search_obs}

Each pulsar candidate was observed using the high time resolution, beam-forming observing modes of LOFAR. Signals from the HBAs of up to 24 LOFAR core stations (longest baseline 3.6\,km) were coherently added in the \textsc{COBALT} correlator and beam former \citep{bmn+18} to form a single, tied-array beam with $\sim3\farcm5$ FWHM for observing frequencies of $110-188$\,MHz. The tied-array beam data were recorded to disc as dual-polarisation, Nyquist-sampled complex voltages for 400 sub-bands of 195.3125\,kHz bandwidth each. The candidate sources were observed twice, at least three days apart, for integration times of 20\,min, except for the faintest sources (ILT\, J104140.11+542419.3 and ILT\,J104937.86$+$582217.6), for which we obtained a single 80\,min integration.

\subsection{Pulsar search pipelines}\label{subsec:search_red}

All of the observations were searched for pulsations using three different strategies, all employing phase-coherent dispersion removal using the GPU-accelerated \textsc{cdmt} software \citep{bph17} to minimize the impact of dispersive smearing. 
These strategies are chosen to retain sensitivity to pulsars over a wide range of spin periods, for example, ranging from PSR\,J0952$-$0607 with $P=1.41$\,ms \citep{bph+17} to PSR\,J0250$+$5854 with $P=23.5$\,s \citep{tbc+18}.
%\LEt{what is ms here? - just checking in case this is repetition?}

The first strategy is optimised for pulsars with short spin periods. Here, the complex voltages were coherently de-dispersed using \textsc{cdmt} to a set of trial DMs in steps of 1\,pc\,cm$^{-3}$, from $\mathrm{DM}=0.5$\,pc\,cm$^{-3}$ up to 79.5\,pc\,cm$^{-3}$ or 159.5\,pc\,cm$^{-3}$ depending on the predicted maximum dispersion measure for the coordinates of each candidate source. Filter-bank files are output with 81.92\,$\upmu$s of time and a 48.83\,kHz frequency resolution. Radio frequency interference (RFI) was masked using the \textsc{rfifind} tool from the \textsc{Presto} software suite \citep{ran01}. The coherently de-dispersed, RFI-cleaned filter banks were then incoherently de-dispersed using the GPU-accelerated \textsc{Dedisp} code \citep{bbbf12} around each of the coherently de-dispersed DM values ($-0.5\leq\Delta\mathrm{DM}\leq0.5$\,pc\,cm$^{-3}$) in steps of 0.002\,pc\,cm$^{-3}$. This approach limits the residual dispersive smearing to less than 0.24\,ms, retaining sensitivity to MSPs with short spin periods. The resulting de-dispersed time series were processed using standard tools from the \textsc{Presto} software suite, which creates the power spectrum, corrects it for red noise, and then uses a GPU-accelerated version of the frequency domain acceleration method \citep{rem02} to search for periodic, and possibly accelerated, signals. The 200 pulsar candidates with the highest signal-to-noise ratio output by \textsc{Presto} were folded for visual inspection. 

The second strategy is a variation of the first, where coherently de-dispersed and RFI-cleaned filter banks were generated at coarser DM steps (2.0\,pc\,cm$^{-3}$) and with coarser time (655.36\,$\upmu$s) and frequency (97.66\,kHz) resolution. This strategy is optimised for pulsars with normal spin periods ($\sim1$\,s), which may not have passed the selection threshold of the first search strategy. These filter-bank files were incoherently de-dispersed in steps of 0.005\,pc\,cm$^{-3}$ around the coherent DM of each filter bank ($-2\leq\Delta\mathrm{DM}\leq2$\,pc\,cm$^{-3}$), providing a maximum dispersive smearing of 1.1\,ms. These time series were searched for pulsations, but not for acceleration. Again, the 200 pulsar candidates with the highest signal-to-noise ratio output by \textsc{Presto} were folded for visual inspection.
%\LEt{missing noun? }

We refer to these first and second search strategies as the `fast' and `slow' pipelines (i.e.\ more suitable to finding recycled MSPs and non-recycled pulsars), and to the filter-bank files produced as having `fine' and `coarse' time/frequency resolution, respectively. 

The third strategy used a fast-folding algorithm (FFA) to search for long period pulsars. Here, the complex voltage data were coherently de-dispersed to a DM of 20\,pc\,cm$^{-3}$ to produce a single filter-bank file with 655.36\,$\upmu$s time resolution and 97.66\,kHz frequency resolution. Next, using incoherent de-dispersion, de-dispersed time series were generated at steps of 0.05\,pc\,cm$^{-3}$ up to a maximum DM of 80\,pc\,cm$^{-3}$ (160\,pc\,cm$^{-3}$ for ILT\,J042737.00$+$684537.0). The \textsc{riptide}\footnote{\url{http://bitbucket.org/vmorello/riptide/}} software \citep{mbs+20} was used to perform FFA searches over four spin-period ranges: 0.3--0.75\,s, 0.75--2.0\,s, 2.0--5.0\,s, and 5.0\,s--2.0\,min. Diagnostic plots were generated for up to 100 of the highest signal-to-noise ratio candidates for visual inspection.

\subsection{Follow-up observations}\label{subsec:tim}

Pulsar detections were confirmed using both LOFAR observations described in \S \ref{subsec:search_obs}. 
Once confirmed, new pulsars were observed using LOFAR at a roughly monthly cadence for at least one year for the purpose of determining a phase-coherent pulsar-timing model describing the rotational and astrometric parameters, and, for those pulsars in a binary system, the binary parameters. 
The confirmation and timing observations are reduced using the LOFAR PULsar Pipeline (PULP) for standard observing modes including complex voltage data \citep[e.g.][]{sha+11,Kondratiev+2016}, which includes coherent de-dispersion and folding using \textsc{dspsr} \citep{sb10} and outputs files in PSRFITS format. For further analysis, we used \textsc{psrchive} \citep{hsm04} and \textsc{tempo2} \citep{hem06,ehm06}. Observations of some new discoveries were also obtained with the 76\,m Lovell telescope at Jodrell Bank over a 400\,MHz band centred at 1532\,MHz.

The DMs were measured using the \textsc{pdmp} routine in the PSRCHIVE software suite \citep{hsm04} and refined using \textsc{tempo2} \citep{hem06}, where possible. 
To measure the RMs, we used the technique of RM synthesis \citep{Burn1966,Brentjens+2005}, and the associated deconvolution procedure RM CLEAN \citep{Heald+2009_RMCLEAN}, using a publicly available Python package\footnote{\url{http://github.com/gheald/RMtoolkit}} and following the procedure used in \citet{sbg+19}. 
The frequency range ($110-188$\,MHz) and channel bandwidth (195\,kHz) provide a nominal resolution in Faraday depth space (i.e. FWHM of the RMSF) of 0.8\,rad\,m$^{-2}$, and maximum observable $\mathrm{|RMs|=65}$ and 162\,rad\,m$^{-2}$ \citep[at full and 50\% sensitivity, respectively;][]{Brentjens+2005}.

The data were corrected for the DMs and RMs (where significantly measured), before constructing the time- and frequency-averaged polarisation pulse profiles.
The pulse-integrated fractional polarisation measurements were obtained using the time- and frequency-averaged polarisation pulse profiles, as the sum of the linearly/circularly polarised flux density divided by the total intensity flux density over each significant ($>3\times$rms) on-pulse phase bin \citep[e.g.][]{Noutsos+2015}. 

The ionospheric Faraday rotation contribution to the RM measured was computed and subtracted for each observing epoch using the publicly available software \textsc{ionFR}\footnote{\url{http://github.com/csobey/ionFR}} \citep{Sotomayor+2013}, with inputs from the International GNSS Service global ionospheric total electron content maps\footnote{\url{http://cddis.nasa.gov/archive/gps/products/ionex}} \citep[e.g.][]{Hernandez+2009,NOLL2010} and the International Geomagnetic Reference Field\footnote{\url{http://www.ngdc.noaa.gov/IAGA/vmod/igrf.html}} \citep[IGRF-13;][]{Thebault+2015}.

\section{Results}\label{sec:results}

We discovered two new radio pulsars, PSRs\,J1049$+$5822 and J1602$+$3901, at the locations of the linearly and circularly polarised sources ILT\,J104937.86+582217.6 and ILT\,J160218.84$+$390101.8, respectively, which is summarised in Table~\ref{tab:pulsars}. These sources are further discussed below.

In addition, we also detected a further five known radio pulsars, summarised in Table~\ref{tab:pulsars}, which were not among the over 40 known pulsars detected as polarised sources in LoTSS and excluded from the candidate list. 
These pulsars were previously discovered using large, sensitive antennas operating between 111 and 430\,MHz.
While these detected pulsars were known at the time of the TULIPP pulsar candidate selection from the LoTSS polarised sources, (sub-)arcsecond pulsar timing positions for these sources were not available in the literature.
Instead, their positions were known to the nearest minute in right ascension and degree in declination, meaning it was not, a priori, clear if a source matched the LoTSS pulsar candidate (with a positional accuracy at the arcsecond level). Only after the spin period and DM were determined from the LOFAR observations was it possible to unambiguously associate the signal with a previously known pulsar.

\begin{table*}
  \centering
  \footnotesize
  \caption{Pulsars discovered or independently re-discovered (detected) in the TULIPP survey.}
  \label{tab:pulsars}
  \begin{tabular}{llrrrrl}
    \hline\hline
    LoTSS source & Pulsar name & \multicolumn{1}{c}{$P$} & \multicolumn{1}{c}{$p$} & \multicolumn{1}{c}{$\mathrm{DM}$} & \multicolumn{1}{c}{RM} & Discovery notes \\
     & & \multicolumn{1}{c}{(s)} & \multicolumn{1}{c}{(\%)} & \multicolumn{1}{c}{(pc\,cm$^{-3}$)} & \multicolumn{1}{c}{(rad\,m$^{-2}$)} & \\
    \hline
    Discoveries &  &   &  & &   & \\
    \hline
\object{ILT\,J104937.86+582217.6} & \object{PSR\,J1049+5822} & 0.7276 &  61$^{L}$ &  12.364(1) & 3.9(1) & TULIPP (this work) \\
\object{ILT\,J160218.82$+$390101.5} & \object{PSR\,J1602+3901} & 0.003708 &  5$^{|V|}$ &  17.2590(2) & -- & TULIPP (this work) \\
    \hline
    Detections &  &   &  &  &  & \\
    \hline
\object{ILT\,J022042.11$+$362655.7} & \object{PSR\,J0220+3626} & 1.029 & $4^{|V|}$ &  45.422(5) & -- & LPA \citep{Tyul'bashev+2016} \\
\object{ILT\,J074212.22$+$411015.0} & \object{PSR\,J0742+4110} & 0.003139 & $-$ &  20.814(1) & -- & GBNCC \citep{GBNCC} \\ 
\object{ILT\,J163035.93+355042.5} & \object{PSR\,J1630+3550} & 0.003229 &  28$^{L}$ &  17.45494(8) & 8.5(1) & AO327 \citep{dsm+13}\\
\object{ILT\,J221227.42$+$245034.4} & \object{PSR\,J2212+2450} & 0.003907 & $-$ &  25.213(1) & -- & AO327 \citep{dsm+13} \\
\object{ILT\,J222741.76$+$303820.8} & \object{PSR\,J2227+3038} & 0.8424 &  23$^{L}$ $10^{|V|}$ & 19.961(6)  & $-59.4$(1) & Arecibo \citep{Camilo+1996} \\
    \hline
  \end{tabular}
  \tablefoot{Pulse period is presented to four significant figures. The fractional polarisation, $p$, is given for sources that have sufficient a signal-to-noise ratios to be measured using the beam-formed observations, where $^{L}$ denotes linear and $^{|V|}$ absolute circular polarisations (the systematic uncertainties are expected to be $\approx5-10$\%; \citealt{Noutsos+2015}). DM and RM measurements using the beam-formed observations are also provided. These are the mean values obtained from the beam-formed data.}
\end{table*}

\subsection{PSR\,J1049$+$5822}

\object{PSR\,J1049$+$5822} was discovered using the FFT-based periodicity search in the slow pulsar search pipeline in the 80\,min observation of the linearly polarised source ILT\,J104937.86+582217.6.
The FFA algorithm also detected this pulsar. 
The pulsar has a spin period $P=0.7276$\,s at a dispersion measure of $\mathrm{DM}=12.364$\,pc\,cm$^{-3}$. 
For this line of sight and DM, the NE2001 \citep{cl02} and YMW16 \citep{ymw17} Galactic electron density models predict distances of 0.56 and 1.1\,kpc, respectively

\begin{figure}
  \centering
    \includegraphics[height=0.2\textheight]{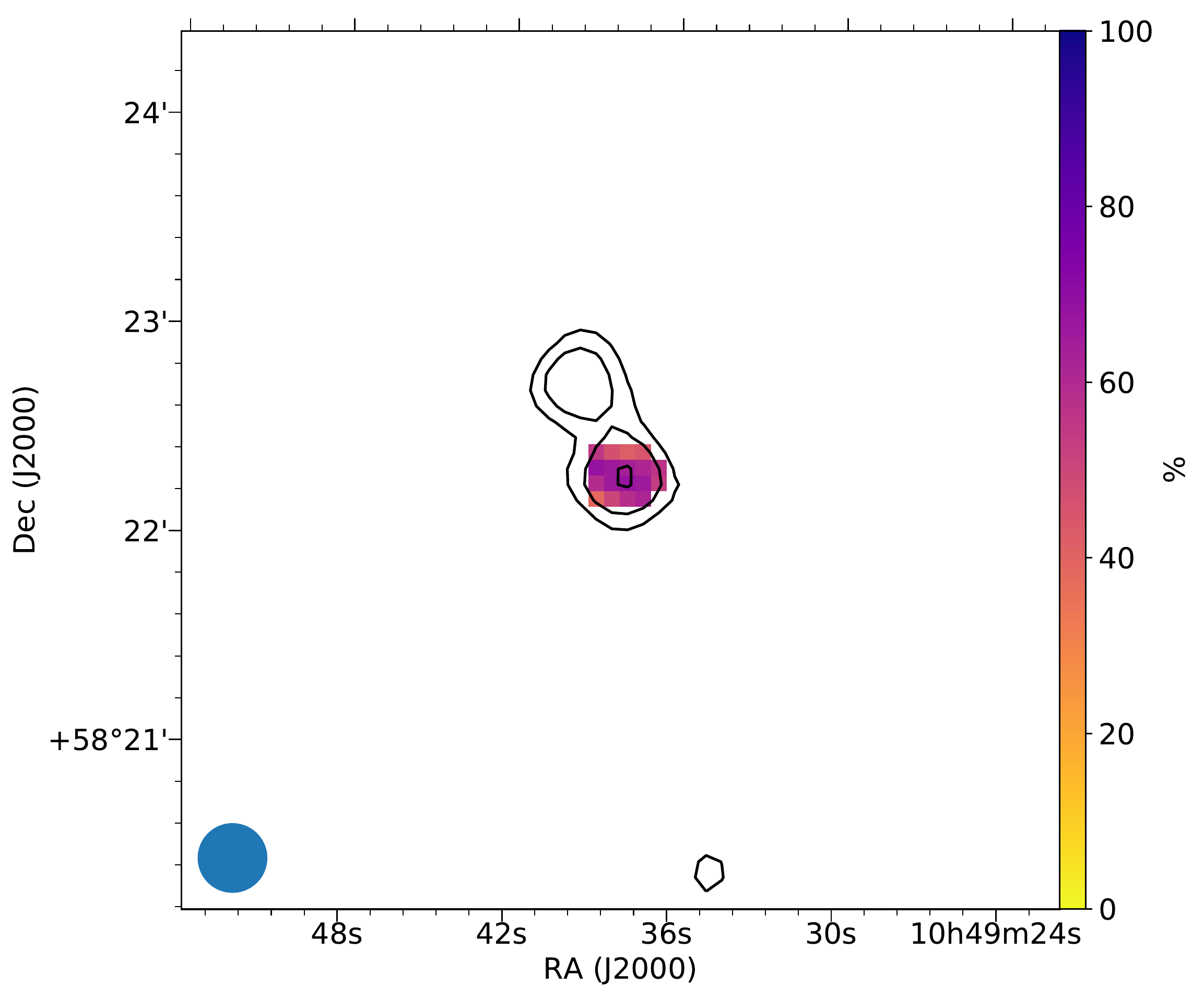}

        \includegraphics[height=0.19\textheight]{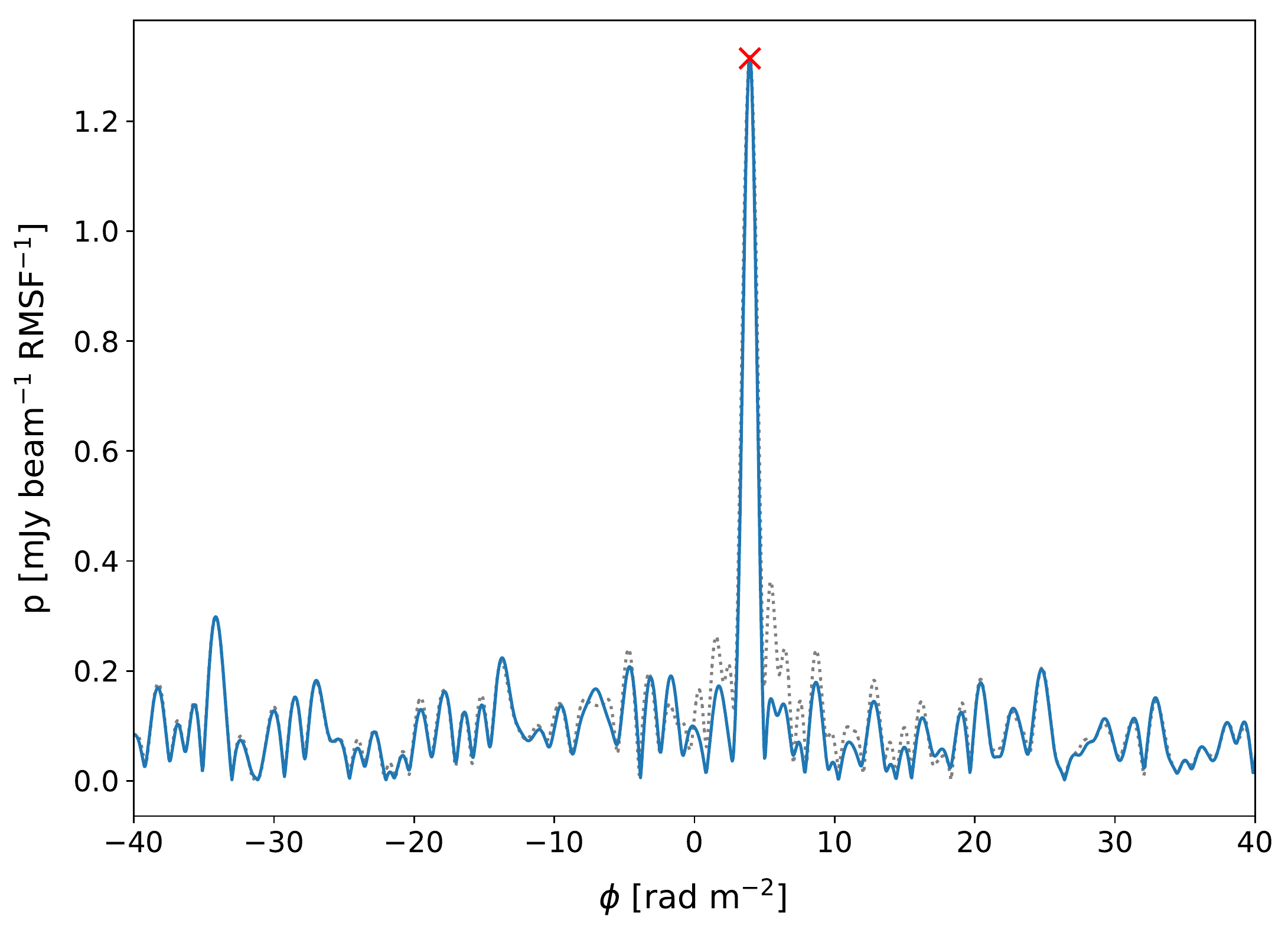}

    \includegraphics[height=0.21\textheight]{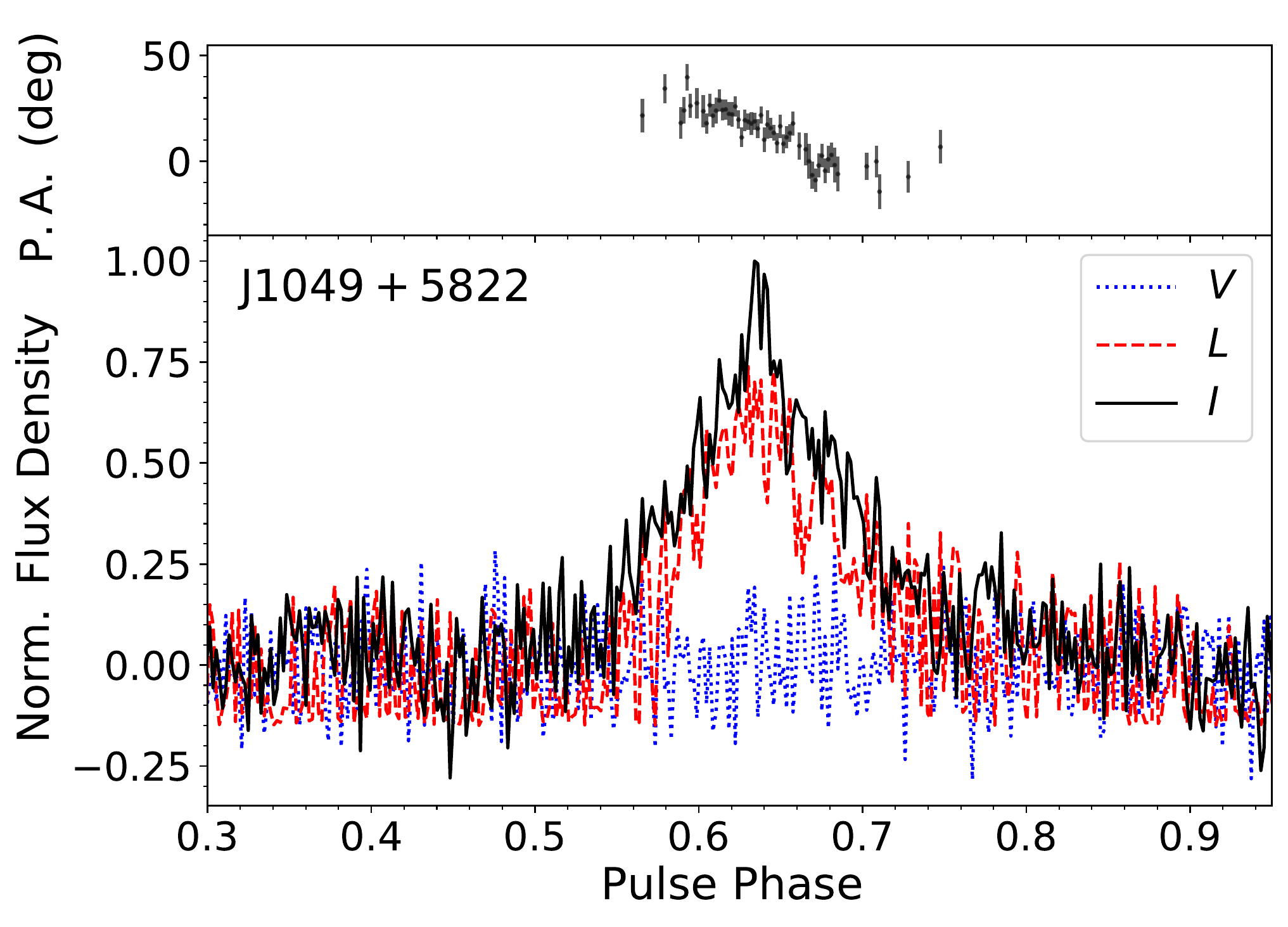}

        \includegraphics[height=0.19\textheight]{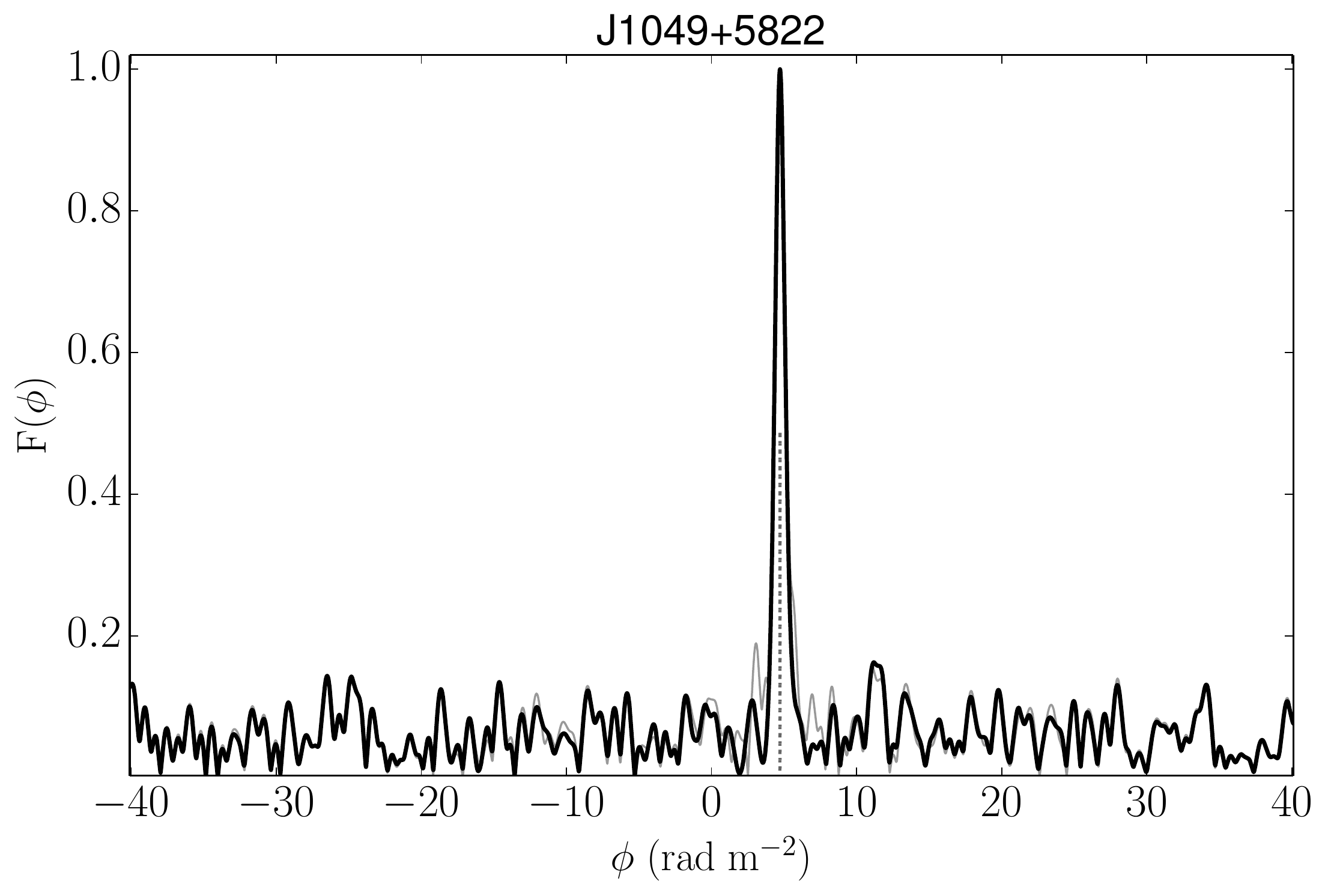}

    \caption{Properties of ILT\,J104937.86+582217.6/PSR\,J1049+5822 discovered in the TULIPP survey, where the panels show the following (top to bottom). 1) LoTSS image at a centre frequency of 144\,MHz with a $20\arcsec$ resolution, contours show the total intensity starting at 0.5\,mJy\,beam$^{-1}$ and increase by factors of 2, and colours show fractional linear polarisation with peak of 63.5\%; the beam size is shown at the lower left corner. The extension to the north is due to another unrelated radio source in LoTSS. 2) Faraday spectrum (original - grey dashed line; RM CLEANed - blue line) computed from \textsc{RM Grid} processing of the LoTSS polarisation image data. The red cross shows the peak corresponding to the $\mathrm{RM}=4.11$\,rad\,m$^{-2}$. 3) Folded polarisation pulse profile, showing the normalised flux density (arbitrary units) for total intensity (Stokes $I$, black line), linear polarisation ($L$, red dashed line), and circular polarisation ($V$, blue dotted line), and the polarisation position angle (arbitrary average offset from $0\degr$). 4) Faraday spectrum (original - grey line; RM CLEANed - black line) computed using PSR\,J1049$+$5822 pulsar observations. The grey dashed line shows the RM CLEAN component at $4.732\pm0.016$\,rad\,m$^{-2}$, prior to the ionospheric RM correction.}
    \label{fig:J1049+5822}
\end{figure}

Figure~\ref{fig:J1049+5822} presents the polarisation data for PSR\,J1049$+$5822, showing the LoTSS image from interferometric LOFAR observations, the polarised pulse profile using the beam-formed LOFAR observations, as well as the Faraday spectra obtained from both of these observations. 
The RMs and the linear polarisation fractions measured using both the LoTSS and beam-formed data show good agreement (see Tables~\ref{tab:sources} and \ref{tab:pulsars}), confirming that this pulsar is indeed the LoTSS source. 
The RM measured towards the pulsar is also $\sim$40\% of the maximum RM expected from the entire Galactic line of sight (see Table~\ref{tab:sources}). 

\begin{figure}
        \includegraphics[width=\columnwidth]{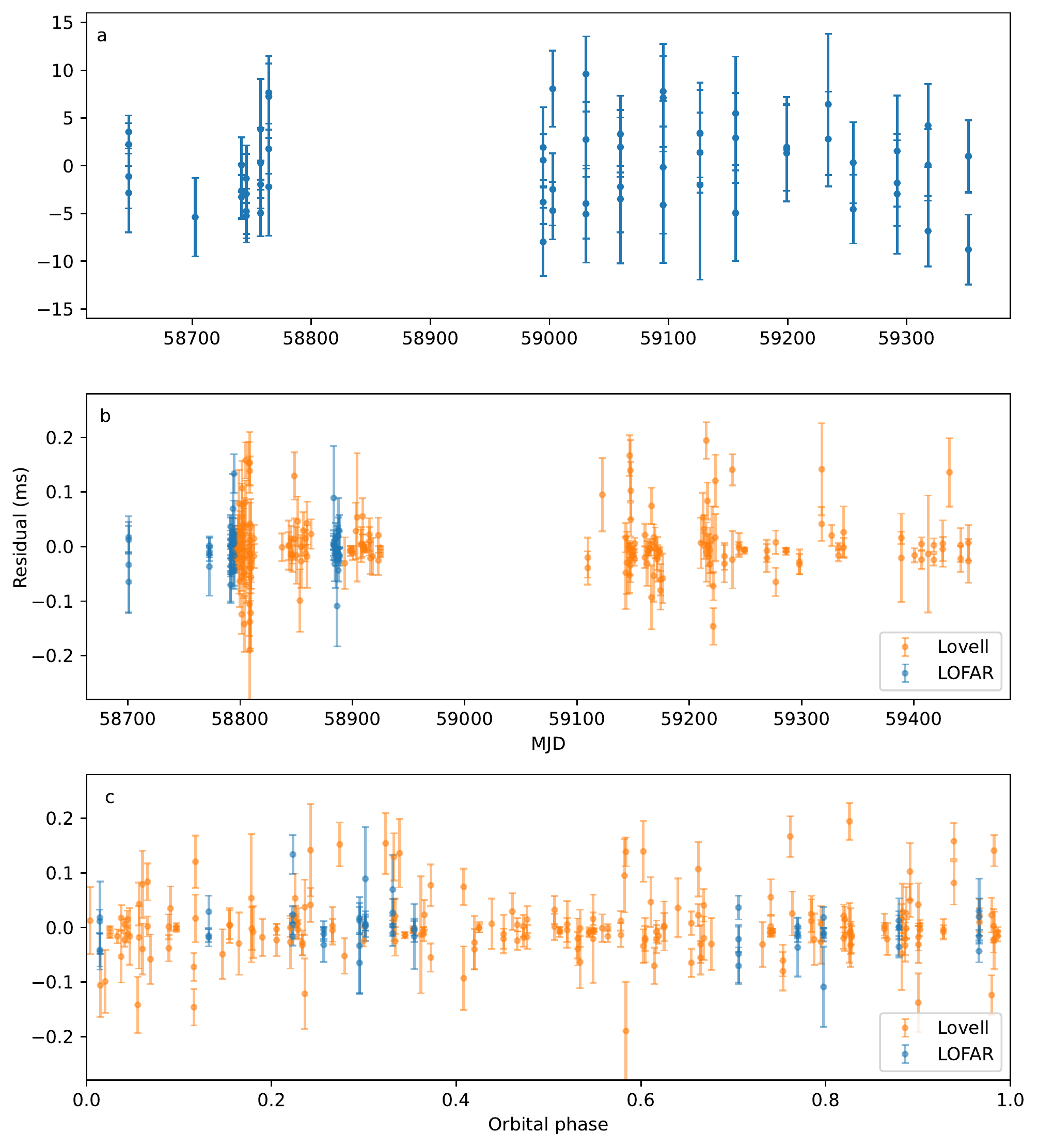}
        
    \caption{Pulsar timing residuals from the timing models in Table\,\ref{tab:tim} for PSR\,J1049+5822 (a) and PSR\,J1630+3550 (b) as a function of time, and PSR\,J1630+3550 as a function of orbital phase (c).}
    \label{fig:J1049+5822_tim}
\end{figure}

We obtained a pulsar timing solution based on LOFAR observations spanning a period of 1.9\,yr, Table~\ref{tab:tim} provides a summary of PSR\,J1049$+$5822's properties based on pulsar timing, and Figure~\ref{fig:J1049+5822_tim} shows the pulsar timing residuals, output from the \textsc{Tempo2} software.   
The timing solution constrains the spin-period derivative to $\dot{P}=3.905(3)\times10^{-15}$\,s\,s$^{-1}$ and provides an arcsecond pulsar-timing position. The pulsar-timing position we measured is in good agreement with the position determined from the LoTSS images, with an offset of $\Delta \alpha=-0\farcs3\pm0\farcs7$ and $\Delta \delta=0\farcs5\pm0\farcs9$. 

\subsection{PSR\,J1602$+$3901}

\object{PSR\,J1602$+$3901} was discovered using the first pulsar search pipeline strategy, applied to a 20\,min observation of the circularly polarised source ILT\,J160218.84$+$390101.8.
The pulsar has a spin period of $P=3.71$\,ms at a dispersion measure of $\mathrm{DM}=17.259$\,pc\,cm$^{-3}$. For this line of sight and DM, the NE2001 and YMW16 Galactic electron density models predict distances of 1.2 and 1.6\,kpc, respectively. 

\begin{figure}
  \centering
        \includegraphics[width=\columnwidth]{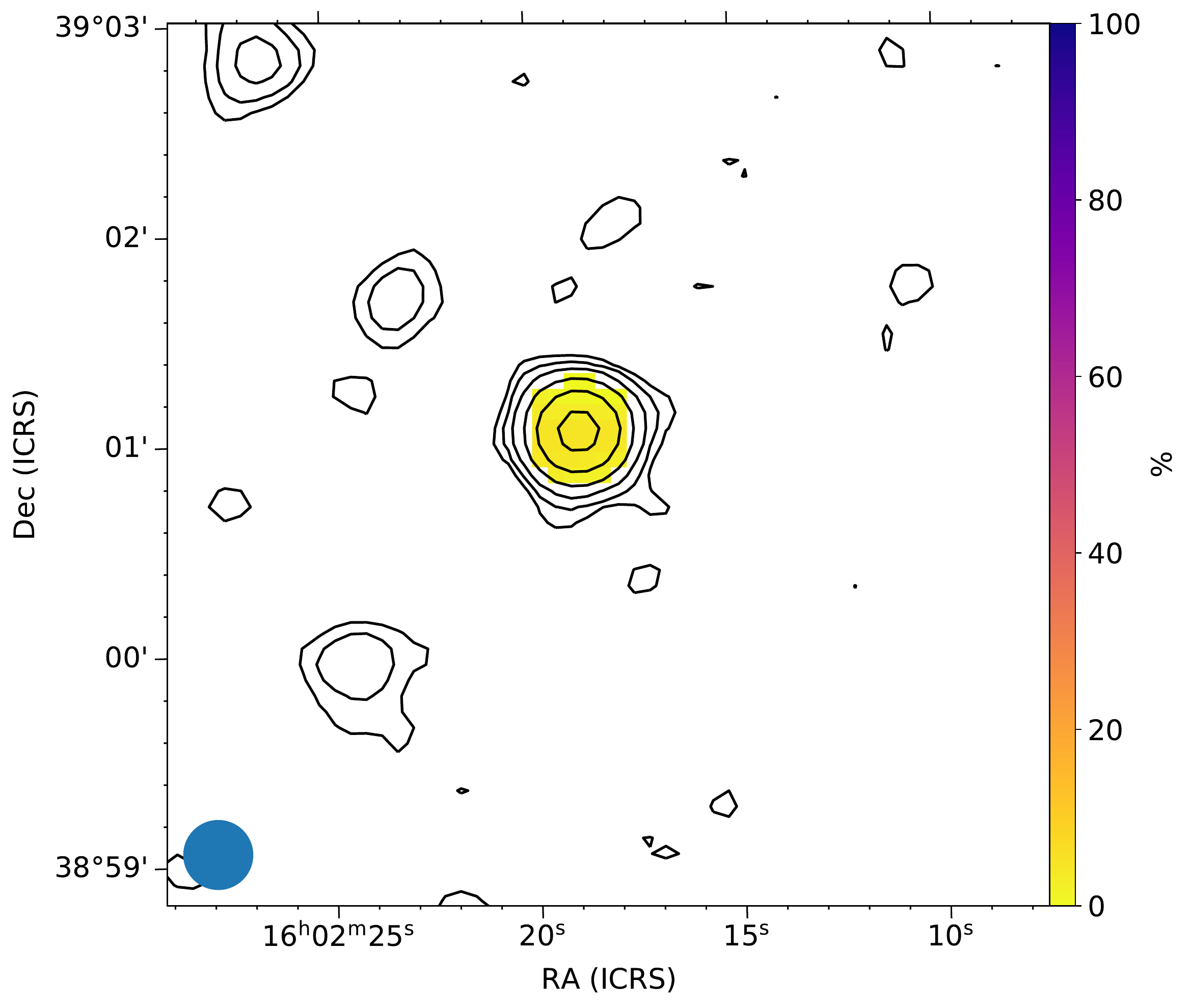}
        
\includegraphics[width=0.45\textwidth]{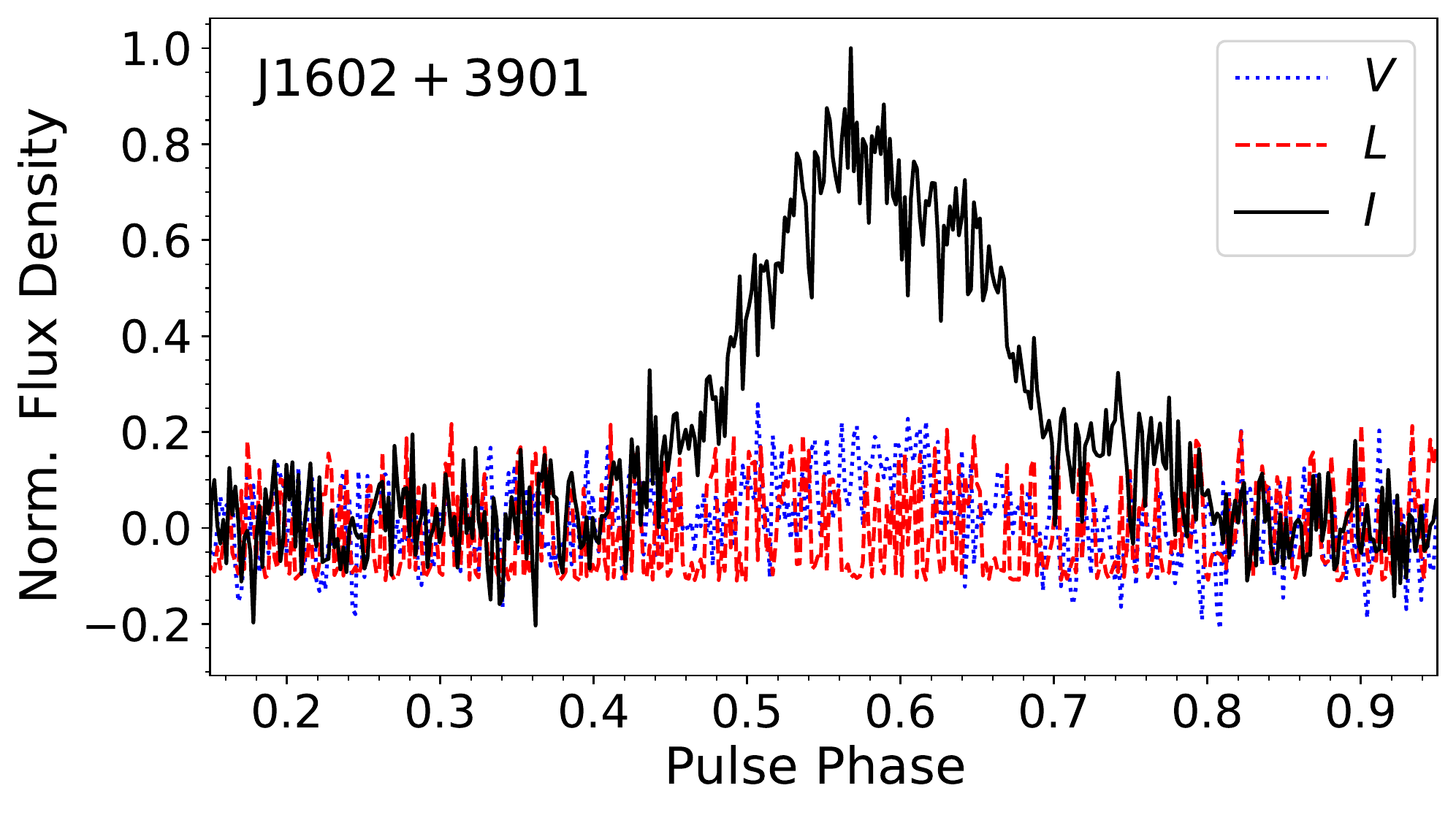}

    \caption{Properties of ILT\,J160218.82$+$390101.5/PSR\,J1602$+$3901 discovered in the TULIPP survey. The top panel shows LoTSS image at a centre frequency of 144\,MHz with a $20\arcsec$ resolution, where contours show total intensity starting at 0.5\,mJy\,beam$^{-1}$ and increase by factors of 2, and colours show fractional circular polarisation with peak of 9\%. The beam size is shown in the lower left corner. The bottom panel shows the polarisation pulse profile. The normalised flux density (arbitrary units) is shown for total intensity (Stokes $I$, black line), linear polarisation ($L$, red dashed line), and circular polarisation ($V$, blue dotted line). Polarisation position angles are not shown, because no linear polarisation was detected above $3\times$ the rms of the off-pulse baseline.  }
    \label{fig:J1602+3901}
\end{figure}

The polarisation properties of PSR\,J1602+3901 are shown in Figure~\ref{fig:J1602+3901}. 
The circular polarisation fractions, measured using the imaging and beam-formed data, agree within the uncertainties (see Tables~\ref{tab:sources} and \ref{tab:pulsars}). 
The beam-formed pulsar data were searched for linear polarisation using RM synthesis, but no significant detection was found.
If the emission from PSR\,J1602$+$3901 were more than a few per cent linearly polarised, we likely would have detected this, as the maximum RM expected from this Galactic line of sight (Table~\ref{tab:sources}) is less than the maximum RM detectable with full sensitivity using these LOFAR data (\S~\ref{subsec:tim}). 

Spin period measurements show that PSR\,J1602+3901 is part of a binary system with an orbital period of $P_\mathrm{b}=6.32$\,d. The binary companion is of low mass ($M_\mathrm{c}\sim0.1$\,M$_\odot$) and likely a Helium core white dwarf. The LoTSS source location is $0\farcs61$ offset from SDSS\,J160218.77+390101.6, an $r=21.0$ object classified as a stellar object \citep{aaa+15sdss} whose colours are a closer match to those of main-sequence stars than those of white dwarfs. The source positions are marginally consistent at $\Delta \alpha=-0\farcs6\pm0\farcs2$, $\Delta \delta=0\farcs2\pm0\farcs2$, suggesting that a chance coincidence cannot be ruled out. Hence, further timing observations are required to obtain a coherent timing solution describing the astrometric, rotational, and binary parameters, and to improve the astrometric matching. The results of these timing observations will be presented in a future publication (Bassa et al., in prep.). 

\subsection{PSR\,J1630$+$3550}

\object{PSR\,J1630$+$3550} is a millisecond pulsar with a $P=3.229$\,ms spin period at a dispersion measure of $\mathrm{DM}=17.455$\,pc\,cm$^{-3}$. This pulsar was originally discovered in the Arecibo drift scan survey at 327\,MHz \citep[AO327;][]{dsm+13}. While the pulsar is not yet published in the ATNF pulsar catalogue, its provisional name (PSR\,J1630+35), spin period, and DM are reported on the AO327 discoveries page\footnote{\url{http://www.naic.edu/~deneva/drift-search/}}.
We independently re-discovered this pulsar in the observations of the linearly polarised source ILT\,J163035.93+355042.5. 
It was detected in both of the FFT-based search pipelines that use fine and coarse time/frequency resolution filter banks. 
The NE2001 and YMW16 Galactic electron density models predict distances of 1.1 and 1.6\,kpc, respectively, for the position and DM of PSR\,J1630$+$3550. 

\begin{figure}
  \centering
    \includegraphics[height=0.2\textheight]{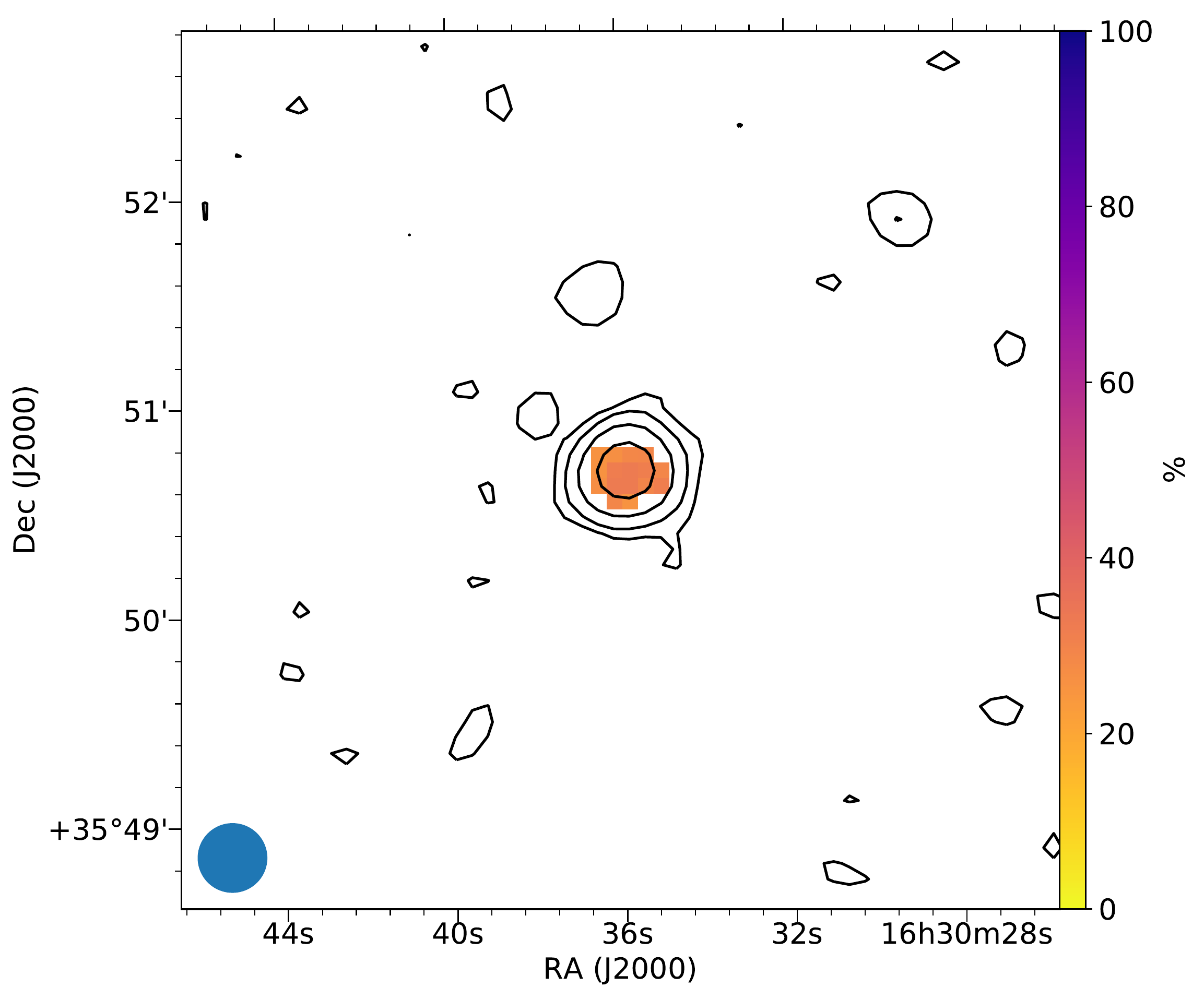}

        \includegraphics[height=0.19\textheight]{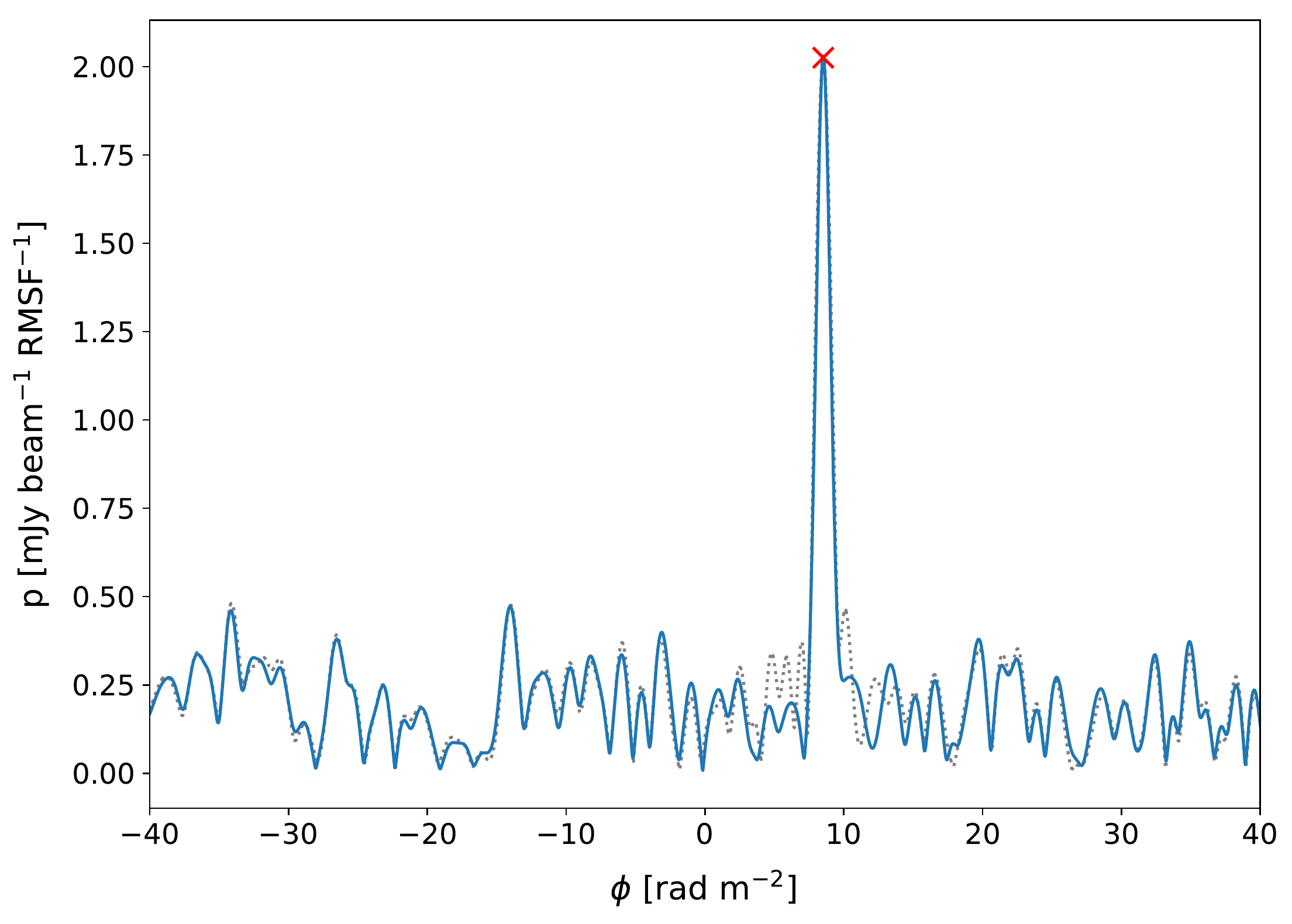}

    \includegraphics[height=0.21\textheight]{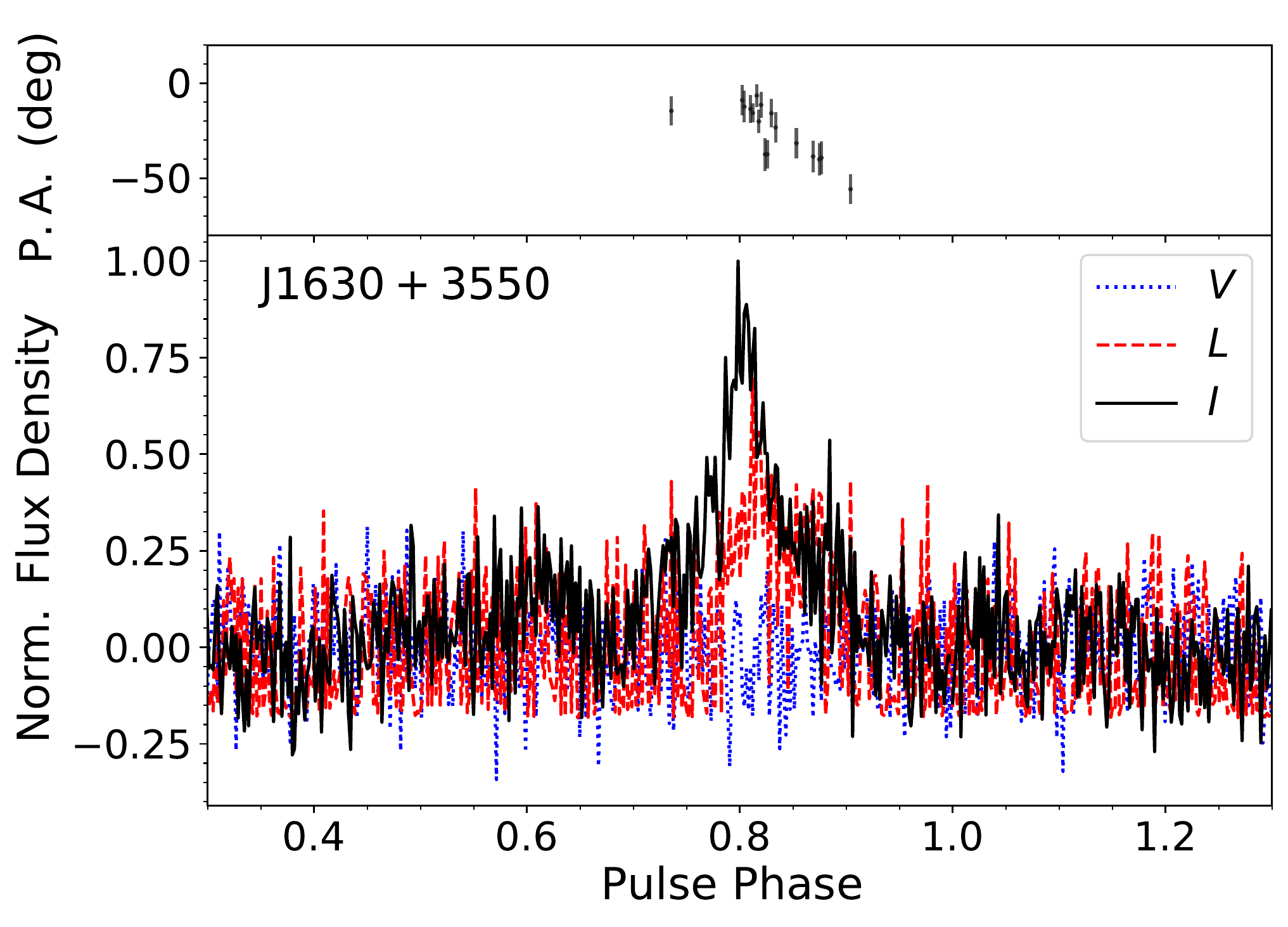}

        \includegraphics[height=0.19\textheight]{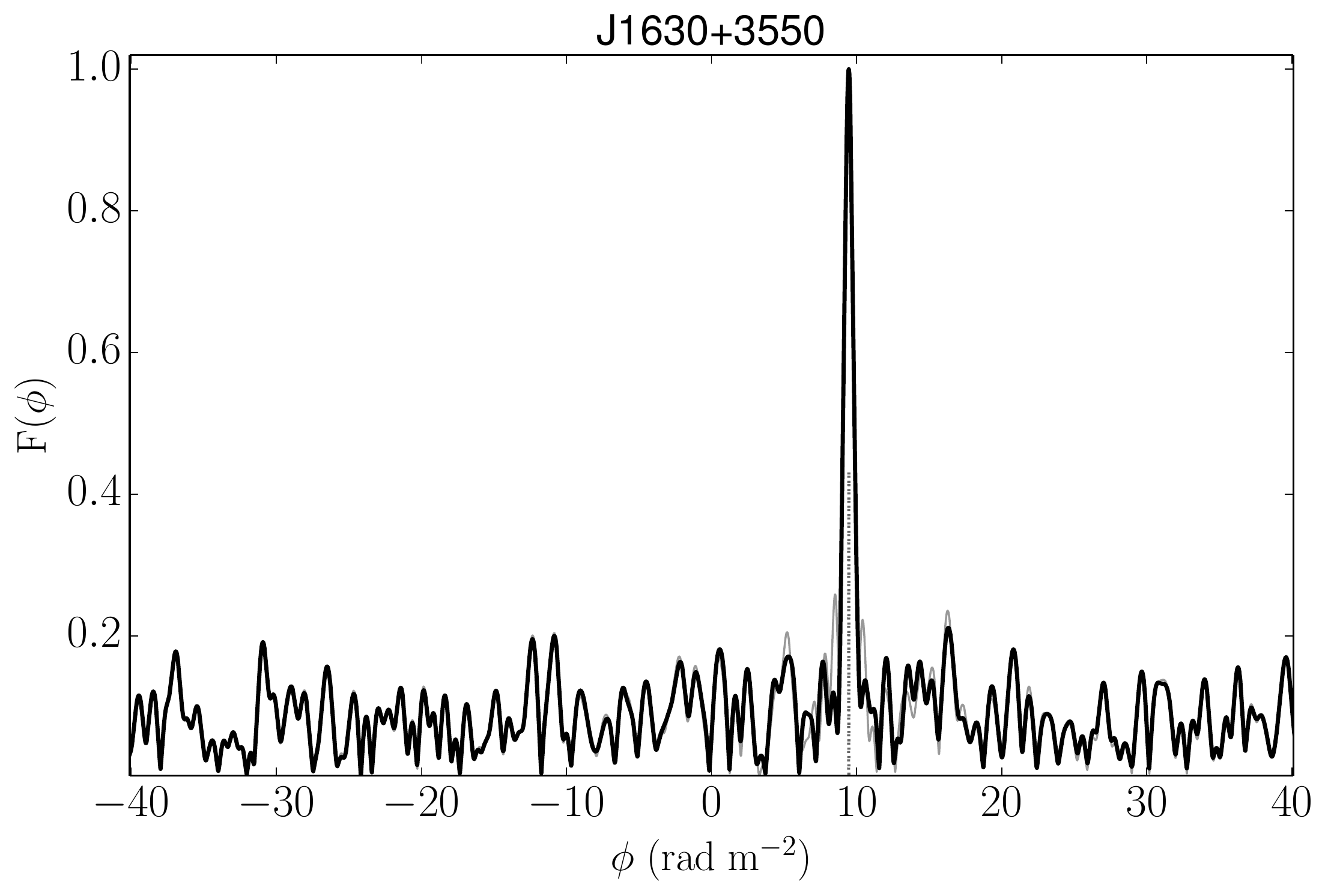}

    \caption{Properties of ILT\,J163035.93+355042.5/PSR\,J1630$+$3550, where the panels show the following (top to bottom). 1) LoTSS image at a centre frequency of 144\,MHz with a 20\arcsec\ resolution, contours show total intensity starting at 0.5\,mJy/beam and increase by factors of 2, and colours show fractional linear polarisation with a peak of 32\%. The beam size is shown at the lower left corner. 2) Faraday spectrum (original - grey dashed line; RM CLEANed - blue line) computed from \textsc{RM Grid} processing using the LoTSS polarisation image data. The red cross shows the peak at 8.65\,rad\,m$^{\rm{-2}}$. 3) Folded polarisation pulse profile from a 20-min observation at 149\,MHz with 78\,MHz bandwidth and 512 pulse profile bins. Lower panel: Normalised flux density (arbitrary units) for total intensity (Stokes $I$, black line), linear polarisation ($L$, red dashed line), and circular polarisation (Stokes $V$, blue dotted line). Upper panel: Polarisation position angle (arbitrary average offset from 0\,deg). 4) Faraday spectrum (original - grey line; RM CLEANed - black line) computed using PSR\,J1630$+$3550 pulsar data. The grey dashed line shows the RM CLEAN component at $9.486\pm0.025$\,rad\,m$^{\rm{-2}}$, prior to the ionospheric RM subtraction.}
    \label{fig:J1630+3550}
\end{figure}

The measured RMs and linear polarisation fractions from the LoTSS imaging and TULIPP beam-formed observations show excellent agreement within the uncertainties, see Tables~\ref{tab:sources} and \ref{tab:pulsars} and Fig.\,\ref{fig:J1630+3550}, confirming that this pulsar is indeed the LoTSS source. The RM measured towards the pulsar is $\sim$65\% of the maximum RM expected from the entire Galactic line of sight.

The discovery and confirmation observations of PSR\,J1630$+$3550 showed a small but non-zero acceleration, indicating that the pulsar could be part of a binary system. 
We used further follow-up pulsar timing observations using both LOFAR and Lovell telescopes and the \textsc{Tempo2} software to confirm the binary nature of this pulsar. Table~\ref{tab:tim} provides the properties of the pulsar and its binary system from timing observations, while Figure~\ref{fig:J1049+5822_tim} shows the pulsar timing residuals. 
PSR\,J1630$+$3550 is orbiting a 0.009\,M$_\odot$ companion in a 7.58\,h orbit, assuming a pulsar mass of 1.4\,$M_\odot$. 
These parameters suggest that PSR\,J1630$+$3550 is part of the \textit{\emph{black-widow}}-class of binary millisecond pulsars, where the companion is irradiated by the energetic wind from the pulsar \citep[e.g.][]{Chen+2013,Polzin+2020}. 
The pulsar timing position we measured (Table~\ref{tab:tim}) and the position determined from the LoTSS images (Table~\ref{tab:sources}) are also in agreement within the uncertainties ($\Delta \alpha=-0\farcs2\pm0\farcs2$, $\Delta \delta=0\farcs1\pm0\farcs2$). No source is coincident with the position of the pulsar in the Pan-STARRS1 survey images \citep{cmm+16}.

\begin{table*}
  \centering
  \caption{Parameters for PSR\,J1049+5822 and PSR\,J1630+3550 based on pulsar timing.}
  \label{tab:tim}
  \begin{tabular}{lrr}
    \hline\hline
    Pulsar name & J1049+5822  & J1630+3550 \\
    \hline
    MJD range  & 58646.7---59351.8 & 58700.7---59448.5 \\ 
    Data span (yr) & 1.93 & 2.05 \\ 
    Number of TOAs & 62 & 242 \\
    RMS timing residual ($\mu$s) & 4080 & 20.1 \\
    Weighted fit & Y & Y \\ 
    Reduced $\chi^2$ value & 1.5  & 2.1 \\
    Reference epoch (MJD) & 59000 & 59000 \\
    Clock correction procedure & TT(TAI) & TT(TAI) \\
    Solar system ephemeris model & DE436 & DE436 \\
    Binary model & -- & ELL1 \\
    \hline
    \multicolumn{3}{c}{Measured parameters} \\ 
    \hline
    Right ascension, $\alpha_\mathrm{J2000}$ &  $10^\mathrm{h}49^\mathrm{m}37\fs90(8)$ & $16^\mathrm{h}30^\mathrm{m}35\fs94666(12)$ \\ 
    Declination, $\delta_\mathrm{J2000}$ & $+58\degr22\arcmin17\farcs1(9)$ & $+35\degr50\arcmin42\farcs433(3)$ \\ 
    Proper motion in $\alpha_\mathrm{J2000}$, $\mu_{\alpha} \cos \delta$ (mas\,yr$^{-1}$) & -- & $-32(3)$ \\ 
    Proper motion in $\delta_\mathrm{J2000}$, $\mu_{\delta}$ (mas\,yr$^{-1}$) & -- & $-8(5)$ \\ 
    Pulse frequency, $\nu$ (s$^{-1}$) & $1.37434776862(6)$ & $309.67840994392(8)$ \\ 
    First derivative of pulse frequency, $\dot{\nu}$ (s$^{-2}$) &  $-7.376(6)\times 10^{-15}$ & $-2.002(10)\times 10^{-15}$ \\ 
    Dispersion measure, DM (pc\,cm$^{-3}$)  & $12.364(11)$ & $17.45494(8)$ \\ 
    Orbital period, $P_\mathrm{b}$ (d) & -- & $0.315863156(6)$ \\ 
    Projected semi-major axis of orbit, $x$ (lt-s) & -- & $0.036128(3)$ \\ 
    Time of ascending node passage, $T_\mathrm{asc}$ (MJD) & -- & $59000.089062(4)$ \\ 
    $\eta = e\cos \omega$ & -- & $-1.5(1.3)\times 10^{-4}$ \\ 
    $\kappa = e\sin \omega$  & -- & $-3.4(14.1)\times 10^{-5}$ \\ 
    \hline
  \end{tabular}
  \tablefoot{Uncertainties in parentheses are the nominal $1\sigma$ \textsc{tempo2} uncertainties in the least significant digits quoted.}
\end{table*}

\subsection{Other known pulsars detected }

We independently re-discovered a further four previously known pulsars: PSRs\,J0220+3626, J0742+4110, J2212+2450, and J2227+3038. 
All were selected as circularly polarised LoTSS sources (see Tables~\ref{tab:sources} and \ref{tab:pulsars}). 

PSR\,J0220+3626 was originally discovered by \citet{Tyul'bashev+2016} using the Pushchino Large Phased Array (LPA) telescope at 111\,MHz, providing a positional accuracy of order $15\arcmin$. Its position, spin period, and DM are consistent with our measurements; though, due to the small receiver bandwidth, \citet{Tyul'bashev+2016} could only constrain the DM to a value between 30 and 50\,pc\,cm$^{-3}$. 
PSR\,J0742+4110 is an MSP that is not yet published, discovered in the Green Bank North Celestial Cap (GBNCC) pulsar survey using the Robert C. Byrd Green Bank Telescope (GBT) at 350\,MHz \citep{GBNCC}. The GBNCC discovery web page\footnote{\url{http://astro.phys.wvu.edu/GBNCC/}} reports it as PSR\,J0740+41, with a spin period of $P\sim3.14$\,ms and $\mathrm{DM}\sim21$\,pc\,cm$^{-3}$, consistently with our measurements. 
Similarly, PSR\,J2212+2450 is an MSP that is not yet published, discovered by the Arecibo AO327 survey \citep{dsm+13}. The AO327 discovery web page reports it as PSR\,J2212+24 with parameters consistent with our measurements ($P=3.91$\,ms and $\mathrm{DM}=25.2$\,pc\,cm$^{-3}$). 
Finally, \citet{Camilo+1996} discovered PSR\,J2227+3038 as PSR\,J2227+30 using Arecibo observations at 430\,MHz taken in 1994. The spin period and DM reported are consistent with our measurements ($P=0.842$\,s and $\mathrm{DM}=33$\,pc\,cm$^{-3}$).

\begin{figure}
  \centering
    \includegraphics[width=0.45\textwidth]{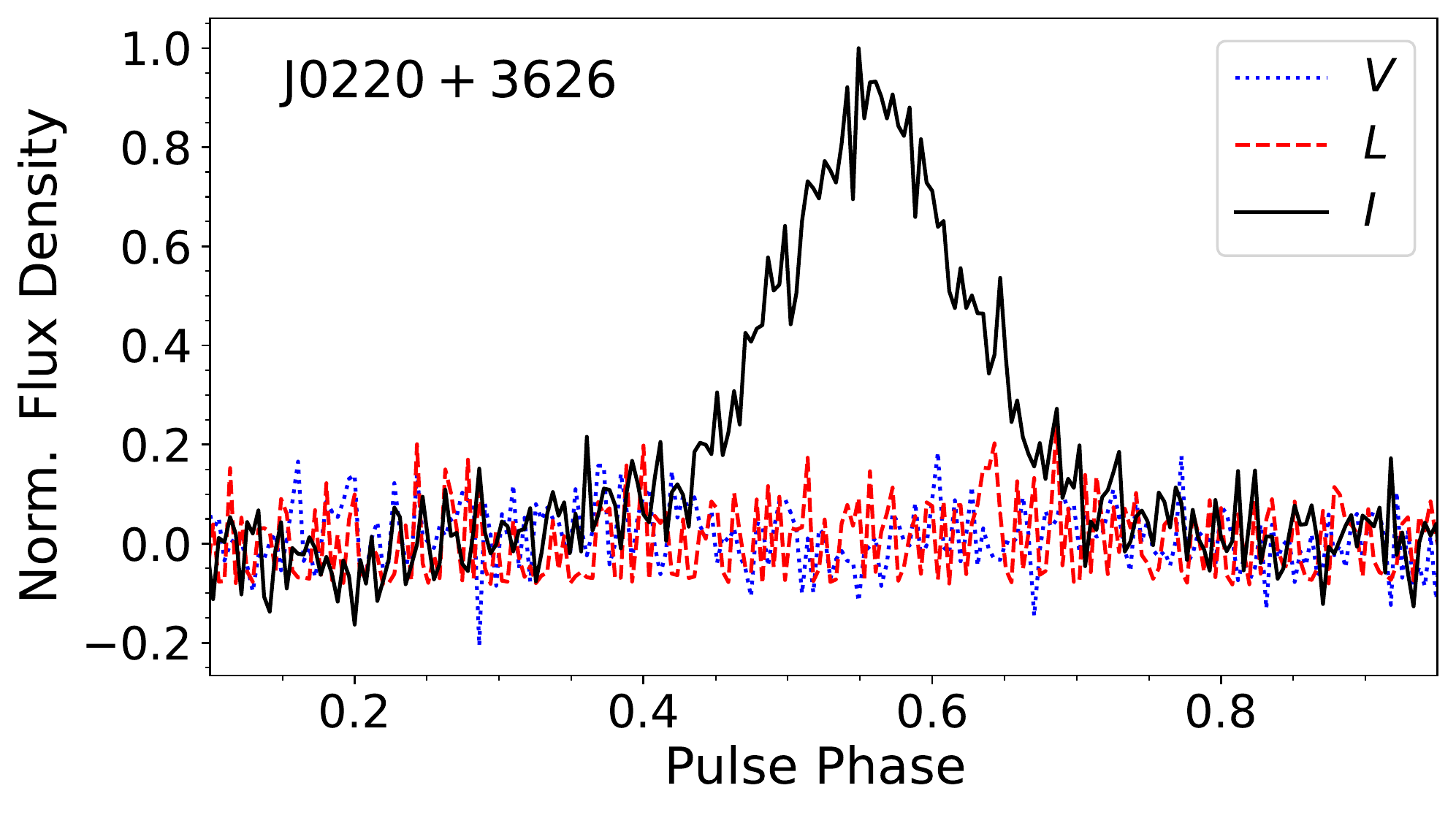}
    
    \includegraphics[width=0.45\textwidth]{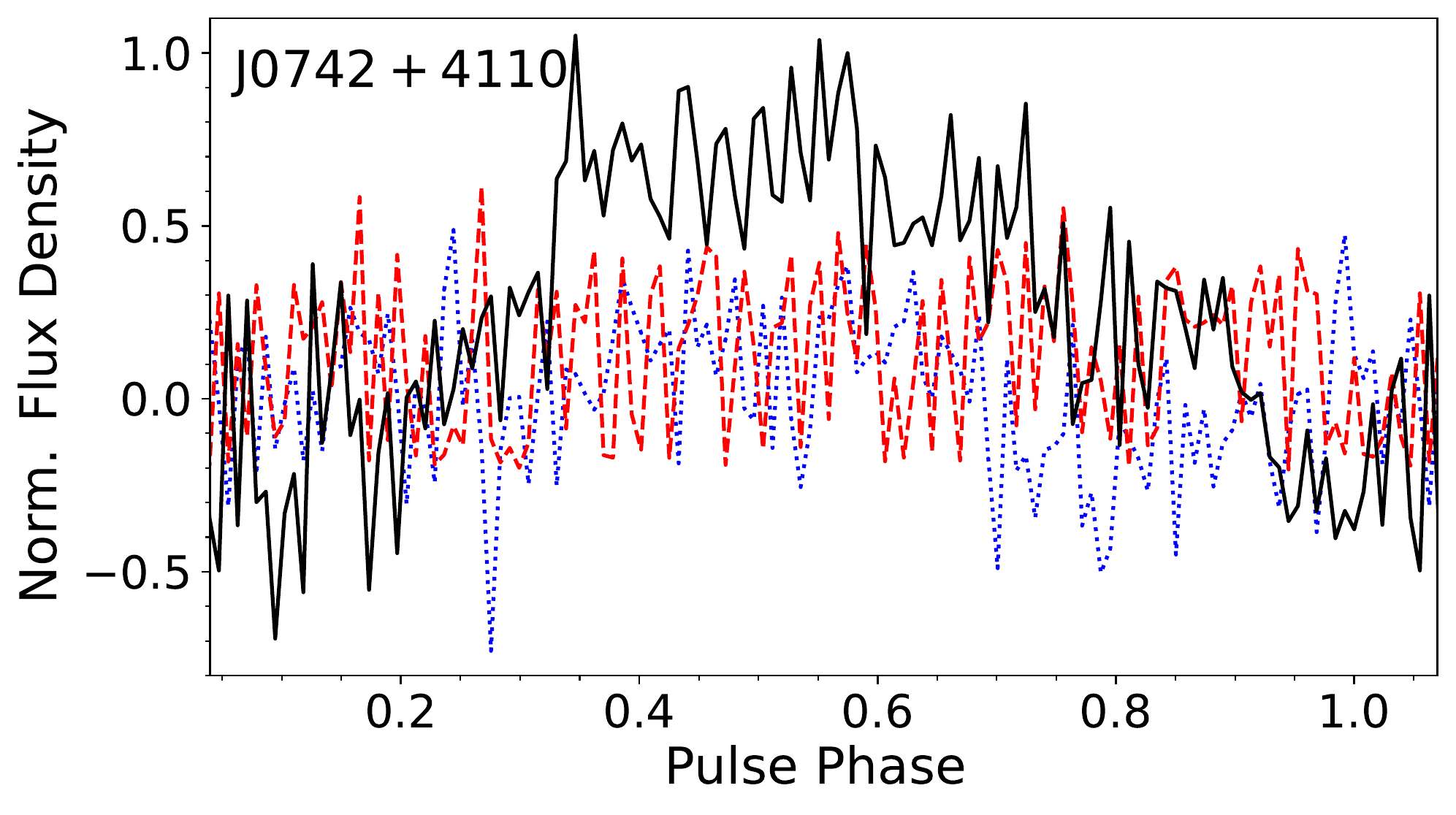}

    \includegraphics[width=0.45\textwidth]{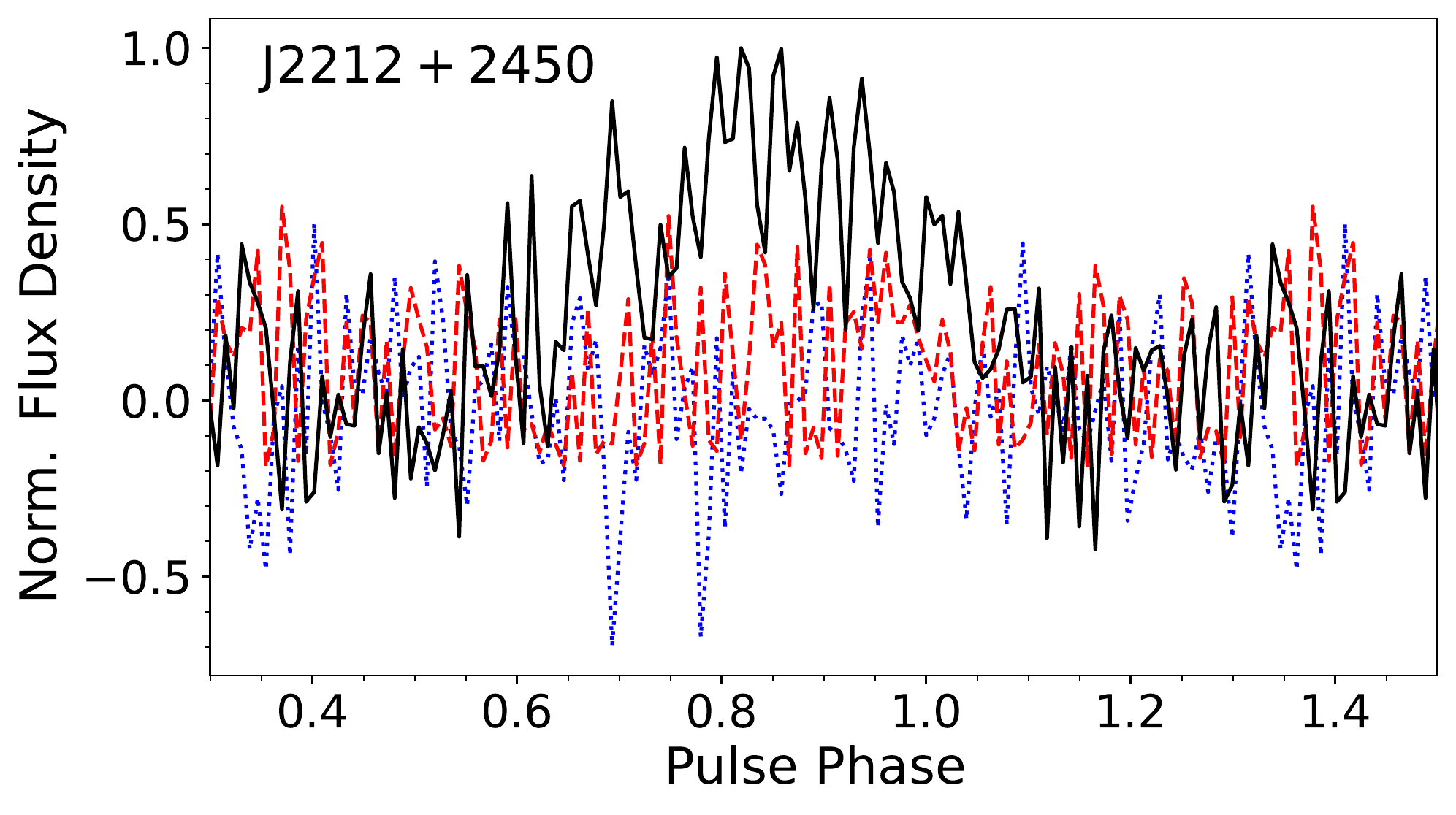}

   \caption{Polarisation pulse profiles for known PSRs\,J0220+3626 (upper), J0742+4110 (middle), and J2212+2450 (lower), respectively, detected in the TULIPP survey. Each was obtained using 20-min observations at a centre frequency of 149\,MHz with a 78\,MHz bandwidth, and 256, 128, and 128 pulse profile bins, respectively. The normalised flux density (arbitrary units) is shown for total intensity (Stokes $I$, black lines), linear polarisation ($L$, red dashed lines), and circular polarisation (Stokes $V$, blue dotted lines). Polarisation position angles are not shown, because no linear polarisation was detected above 3$\times$rms of the off-pulse baseline. }
    \label{fig:other_PSRs}
\end{figure}

\begin{figure}
  \centering
    \includegraphics[width=0.45\textwidth]{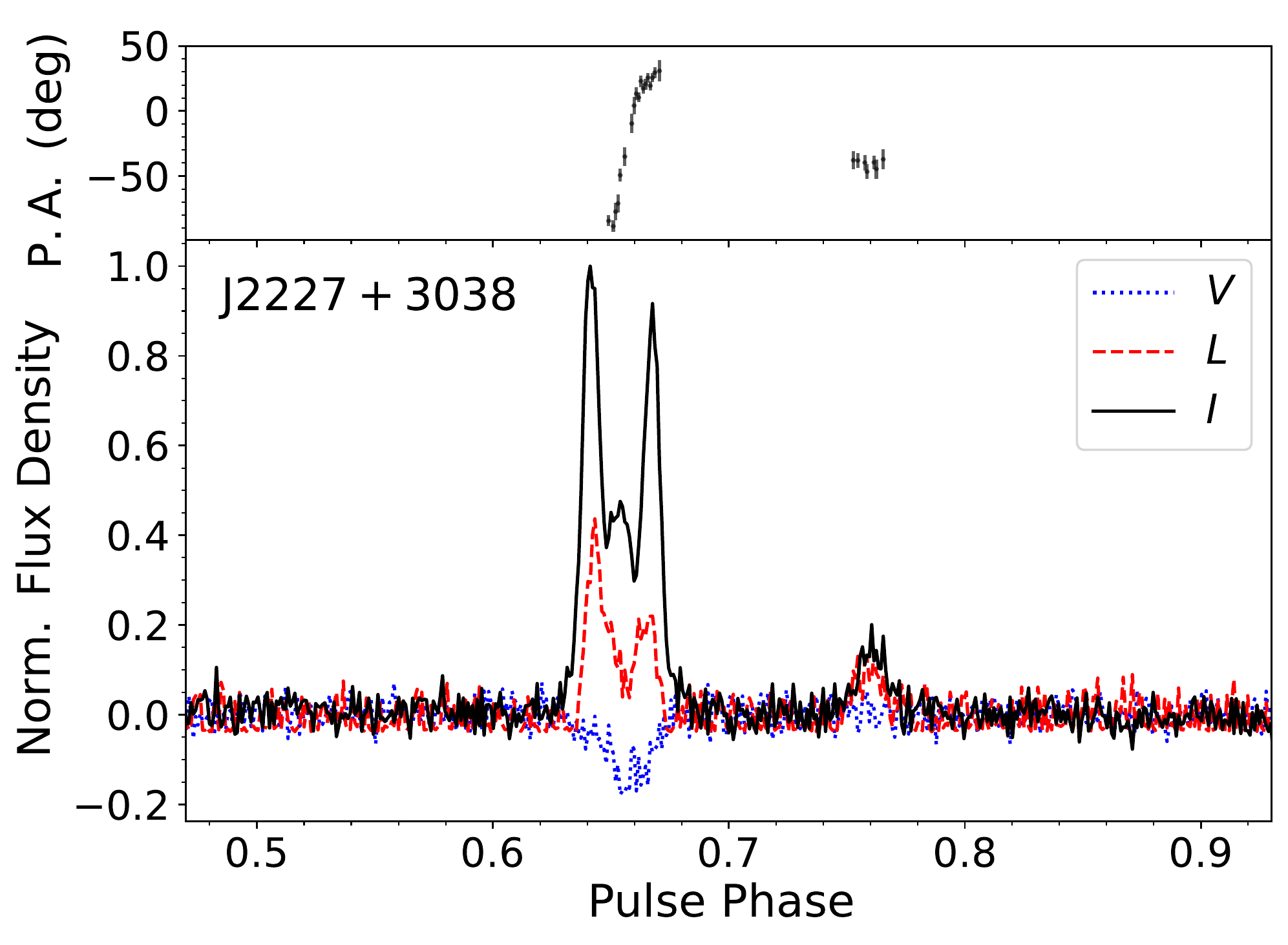}
    
    \includegraphics[width=0.45\textwidth]{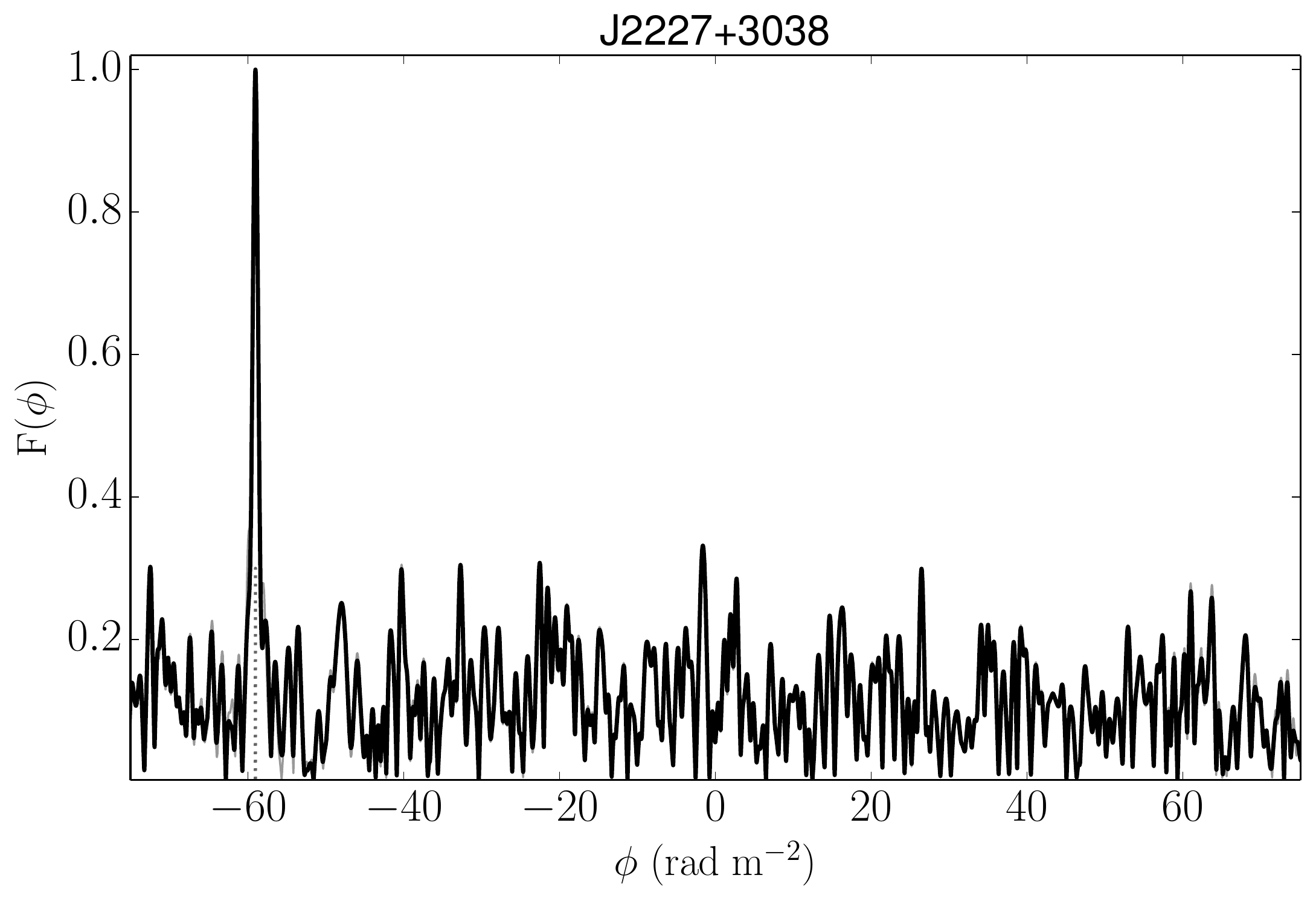}

   \caption{Properties of the known PSR\,J2227$+$3038 detected in the TULIPP survey. Upper: Polarisation pulse profile obtained from a 20-min observation at a centre frequency of 149\,MHz with a 78\,MHz bandwidth and 1024 pulse profile bins, showing the normalised flux density (arbitrary units) for total intensity (Stokes $I$, black line), linear polarisation ($L$, red dashed line), circular polarisation (Stokes $V$, blue dotted line), and polarisation position angle (arbitrary average offset from 0\,deg). Lower: Faraday spectrum (original - grey line; RM CLEANed - black line) computed using PSR\,J2227$+$3038 pulsar data. The grey dashed line shows the RM CLEAN component at $-59.04\pm0.03$\,rad\,m$^{\rm{-2}}$, prior to the ionospheric RM subtraction. }
    \label{fig:J2227+3038}
\end{figure}

The LOFAR polarisation pulse profiles for these four pulsars are shown in Figures~\ref{fig:other_PSRs} and \ref{fig:J2227+3038}. 
The circular polarisation fractions of PSRs\,J0220+3626 and J2227+3038 are consistent within uncertainties between the LoTSS and the pulsar data (Tables~\ref{tab:sources} and \ref{tab:pulsars}). 
No linear polarisation was detected from PSRs\,J0220+3626, J0742+4110, or J2212+2450 after processing the data using RM synthesis. 
Similarly to PSR\,J1602$+$3901, if the emission from PSR\,J0220+3626 were $\gtrsim10$\% linearly polarised, we likely would have detected this, as the maximum RM expected from the Galactic line of sight (Table~\ref{tab:sources}) is less than the maximum RM detectable at 50\% sensitivity with these LOFAR observations (\S~\ref{subsec:tim}).
No polarised emission could be detected from PSRs\,J0742+4110 or J2212+2450 due to the lower signal-to-noise ratio of their average pulse profiles. 

PSR\,J2227+3038 shows relatively high fractional linear polarisation and the RM was significantly detected in both LOFAR pulsar search observations (see Table~\ref{tab:pulsars} and Figure~\ref{fig:J2227+3038}).  
The associated source ILT\,J222741.76+303820.8 in the LoTSS image is weakly detected but is too faint to be included in the \textsc{RM Grid} catalogue with an $S/N<7$.
However, after manual inspection of the data, we measure an $\mathrm{RM}=-58.7 \pm 0.1$\,rad\,m$^{-2}$ and fractional linear polarisation of $\approx6$\%. 
The RMs we measured towards PSR\,J2227+3038 are in good agreement within the uncertainties and with the measurement published in the ATNF pulsar catalogue: $-58.3 \pm 1.9$\,rad\,m$^{-2}$ \citep{Ng+2020}.
PSR\,J2227+3038 shows a steep `S'-shaped curve in the polarisation position angle (P.A.) with a pulse phase across the main pulse with a range $\approx130\degr$ and a highly linearly polarised post-cursor component with approximately flat P.A.s (Figure~\ref{fig:J2227+3038}). 
The P.A. range across the main pulse profile is larger than for PSRs\,J1049+5822 and J1630+3550, the other pulsars detected in linear polarisation, which have relatively flat P.A. swings across their pulse profiles.
The LoTSS data provide the average measurement across the pulse profile, and this P.A. range likely acts to de-polarise the signal, accounting for the lower linear fractional polarisation measured in the \textsc{RM Grid} data (e.g. Posselt et al., in prep.).

\section{Discussion}\label{sec:discussion}

\subsection{Pulsars (re-)discovered in TULIPP }

During the first year of the TULIPP survey, we discovered two new pulsars and detected a further five known pulsars towards either linearly or circularly polarised LoTSS point sources. 
The pulsars presented in this work have been characterised using both LOFAR imaging and beam-forming modes (Tables~\ref{tab:sources} and \ref{tab:pulsars}).  
The pulsars we detected have spin periods between $P=3.1$\,ms and $1.03$\,s, dispersion measures between $\mathrm{DM}=12$\,pc\,cm$^{-3}$ to 45\,pc\,cm$^{-3}$, and 144\,MHz flux densities in the range of $S_\mathrm{144}=2.1$\,mJy to 24\,mJy.
In general, the measured properties using both the imaging and beam-formed data are in agreement, including the sky coordinates (for PSRs\,J1049+5822 and J1630+3550), fractional polarisations, and RMs. 
Although, we do not necessarily expect the fractional polarisation measurements to agree (e.g. PSR\,J2227+3038), because the beam-formed observations provide pulse phase-resolved information, within which the polarised flux densities and P.A.s vary, while the imaging measurements provide the average across the pulse profile. 
We did not measure flux densities using the beam-formed data, because the imaging data are more reliable and provide smaller uncertainties \citep[e.g.][]{Bilous+2016}. 
We also did not present the in-band spectral indices using the imaging or beam-formed data, because the uncertainties are large.
We are collecting multi-wavelength data, which will provide more reliable wide-band radio spectra and will be presented in future work.  

Only one of the pulsars, PSR\,J2227+3038, shows significant emission in both linear and circular polarisations. It also shows the most complex pulse profile, with reasonably narrow triple-peaked main pulse and post-cursor components. The other pulsars generally show a reasonably wide (relatively large duty cycle), single-peaked pulse profile with either significant linear or circular polarisation. This is not unusual for the MSPs, but it is somewhat less common for the non-recycled pulsars PSRs\,J0220+3626 and J1049+5822 \citep[e.g.][]{Kramer+1998,Pilia+2016,Kondratiev+2016}. Furthermore, the pulse profile of PSR\,J0742+4110 shows the largest duty cycle ($\sim$70\%) for this set of pulsars and appears to be quite heavily affected by interstellar scattering. No higher frequency pulse profiles are published to further investigate this.

% Discuss why not discovered previously in old pulsar surveys and LOTAAS?
LOTAAS pointings towards all of the candidate sources in Table~\ref{tab:sources} have already been observed, but no pulsar was identified in the majority of these data. 
However, the non-recycled pulsars PSRs\,J0220+3626 and J2227+3038 (spin periods of $P=1.029$\,s and $P=0.842$\,s) that were detected in this work were also detected in the LOTAAS survey \citep{Sanidas+2019}. % 14.6, 14.5 mJy
These sources have higher flux densities ($S_{144}>15$\,mJy) than PSR\,J1049+5822, the non-recycled pulsar that was discovered in the TULIPP survey. 
PSR\,J1049+5822 lies in the $P-{\rm DM}-S$ parameter space where LOTAAS discovered or detected known pulsars \citep{Sanidas+2019}. 
However, due to the low flux density of $S_{144}=2.0$\,mJy, the integration time (1\,h), and number of LOFAR HBA stations used in the LOTAAS observations (6), the expected signal-to-noise ratio of PSR\,J1049+5822 is $\approx$8, and near the LOTAAS detection significance threshold. Folding and dedispersing the LOTAAS observation containing PSR\,J1049+5822 with the timing solution from Table\,\ref{tab:tim} did not lead to a significant detection.

This work also highlights the potential for discovering MSPs through TULIPP. 
Due to the coarse time resolution of the LOTAAS survey, its sensitivity to detect MSPs was significantly reduced compared to the observational setup of TULIPP.
The six known MSPs that were detected in LOTAAS are relatively bright ($>25$\,mJy) and have relatively long spin periods ($P>4.2$\,ms) and relatively low DMs ($\mathrm{DM}\leq18.3$\,pc\,cm$^{-3}$); the shortest spin period pulsar discovered using LOTAAS is PSR\,J0827+53, a binary MSP with $P=13.5$\,ms and $\mathrm{DM}=23.1$\,pc\,cm$^{\rm{-3}}$ \citep{Sanidas+2019}.
Known non-recycled pulsars, and MSPs with spin periods $2<P<6$\,ms, have been detected in the LoTSS linear polarisation images (O'Sullivan et al., in prep.). 
All of these non-recycled pulsars were also detected in the LOTAAS survey, while 70\% of these MSPs, all with spin periods $P<4.2$\,ms, were not detected \citep{Sanidas+2019}. 
The MSPs PSRs\,J0742+4110, J1630+3550, and J2212+2450 with spin periods $P<3.9$\,ms, which were detected in this work, were not detected in LOTAAS due to its reduced sensitivity to fast-spinning pulsars.

New pulsars discovered using the low-frequency LOFAR observations also provide precise DM measurements, in addition to the precise RM measurements from the LoTSS \textsc{RM Grid}. 
Since pulsar DMs and RMs are an efficient method for estimating the average Galactic magnetic field strength and direction parallel to the line of sight  \citep[e.g.][]{sbg+19}), these provide more data that point towards reconstructing the Galactic magnetic field in 3-D, particularly towards the Galactic halo, which is relatively unexplored \citep[e.g.][]{Terral+2017}. 
 
\subsection{Non-detection of other polarised sources as pulsars}\label{subsec:nondet}

% Discussion on possibilities for other sources being pulsars and not detected as such
Although we have successfully (re-)discovered pulsars in the TULIPP survey using the methods outlined here, there are seven linearly polarised and 16 circularly polarised sources that we have not discovered pulsations towards (Table~\ref{tab:sources}). 
This may be due to several factors that are briefly described below, which include intrinsic emission beam geometry or variability, extrinsic variability and other ionised ISM propagation effects, instantaneous observing sensitivity, or sources that are not pulsars.
Similar discussions of these factors have also been presented in the literature \citep[e.g.][]{Crawford+2000,Crawford+2021,Hyman+2019,Maan+2018}.

%\textbf{Short-binary-period systems and variability:}  
We attempted to minimise effects due to variability in flux density with time by repeating observations of the sources twice. 
This provided the opportunity to sample different interstellar scintillation conditions, emission states (in case of nulling or mode-changing, e.g. \citealt{Wang+2007}), or binary period phases (e.g. black widow/redback binary systems show eclipses during some binary period phases; e.g. \citealt{Camilo+2015}; \citealt{Polzin+2020}). 
Although diffractive scintillation bandwidths are small at these low observing frequencies, and therefore the effect is often averaged out over the large bandwidth, refractive scintillation can cause the flux density to vary by a factor of a few on longer timescales \citep[e.g.][]{Cordes+1998,Archibald+2014,Kumamoto+2021}.
In future, a check could be carried out to ensure that the source is indeed still visible at the time of the beam-formed observations using the interferometric imaging mode simultaneously \citep[e.g.][]{Polzin+2020}. 
If these candidate sources are too accelerated to be detected in the pulsar observations, they may be pulsars in very short binary period systems, which are important for testing general relativity \citep[e.g.][]{Cameron+2018}; yet, the most accelerated systems consisting of two neutron stars are typically found at low Galactic latitude.
If this is the case, different approaches to searching for pulsars in this data may be required, for example, template matching or other novel search methods \citep{Balakrishnan+2021}.
Moreover, observations with telescopes with greater instantaneous sensitivity could also prove fruitful, although this may not be possible until the SKA--Low is constructed (for sources visible in the southern hemisphere).

%\textbf{ISM propagation effects:}
As the impulsive signals from pulsars traverse the ISM, they become dispersed and scattered. 
We can effectively correct for the dispersion in the data that we record up to relatively large dispersion measures.
For example, the LOTAAS survey was able to discover and detect non-recycled pulsars (with $P>1$\,s) up to a maximum DM of 131 and 217\,pc\,cm$^{-3}$, respectively \citep{Sanidas+2019}.  
However, the broadening of pulse profiles due to interstellar scattering in the anisotropic ISM can become severe at low frequencies, and we are not yet able to remove this effect from the data we collect for pulsar searches. 
Intervening structures in the ISM, acting as scattering screens, influence the propagation of pulsar emission \citep[e.g. supernova remnants, H \textsc{II} regions; e.g.][]{Mitra+2003,Johnston+2021,Sobey+2021}.
Interstellar scattering times are proportional to wavelength to the power of $\sim$4 (e.g. 4.4, assuming uniform Kolmogorov turbulence in the ISM, e.g. \citealt{Geyer+2017}), and they are correlated with DMs \citep[e.g.][and references therein]{Oswald+2021}. 
Detecting pulsars in continuum images is not affected by the DMs or scattering times of the pulsars.
For example, the POlarised GLEAM Survey using the Murchison Widefield Array (MWA) at 169--231\,MHz \citep{Riseley+2020} detected bright, polarised pulsars with larger DMs (up to 515\,pc\,cm$^{-3}$) compared to those detected using the MWA's high time resolution mode (up to 158.52\,pc\,cm$^{-3}$; \citealt{Xue+2017}).
If the polarised LoTSS sources are MSPs with large DMs (significant fractions of those predicted using the Galactic electron density models) then conducting search observations at slightly higher observing frequencies, where these ISM propagation effects are less severe, may prove fruitful.
In this case, the spectral index of the sources should also be taken into account \citep[e.g.][]{Camilo+2021}. 
Most of the candidates searched in this work are towards Galactic latitudes of $|b|>13$\degr \ (Table~\ref{tab:sources}), where scattering times are relatively small compared to those towards the Galactic plane \citep[e.g.][]{ymw17}. Candidates that are located closer to the Galactic plane, including ILT\,J003904.45+625711.4 at $|b|=0.12$\degr, may require deeper follow-up observations at higher frequencies. 

%ILT J003904.45+625711.4 and ILT J024831.01+423020.3  -- 4FGL J0039.1+6257 and 4FGL J0248.6+4230
Two TULIPP candidates coincident with $\gamma$-ray sources have indeed been searched for radio pulsations at higher radio frequencies. 
The linearly polarised source ILT\,J024831.01+423020.3, coincident with 4FGL\,J0248.6+4230 \citep{fermi+19}, was (more recently than our TULIPP observations) discovered to be an MSP (PSR\,J0248+4230) using a 30-min Giant Metrewave Radio Telescope (GMRT) observation at 322\,MHz \citep{Bhattacharyya+2021}. 
PSR\,J0248+4230 is an isolated MSP with $P=2.6$\,ms, DM=48.2\,pc\,cm$^{-3}$ (the pulse profile shows significant dispersion smearing at 322\,MHz), a steep spectral index ($\alpha=-3.3$), and $<2\upsigma$ agreement between the sky positions from the LoTSS and GMRT images \citep{Bhattacharyya+2021}.
After we re-examined the \textsc{Presto} outputs from the two LOFAR observations of ILT\,J024831.01+423020.3, we found that PSR\,J0248+4230 was only detected as a $9\upsigma$ candidate (ranked 29 out of 200) in one observation.
The pulse profile is visibly affected by interstellar scattering and is only detectable at the highest frequencies of the LOFAR band ($\nu>140$\,MHz). 
This candidate was not selected for further follow-up observations due to the low significance of the detection for a source with a total flux density of 9\,mJy, combined with its absence from the second observation, which can be relaxed in future analyses.
Additionally, the circularly polarised source ILT\,J003904.45+625711.4 is coincident with 3FGL\,J0039.3+6256 (4FGL\,J0039.1+6257; \citealt{fermi+19}) and was predicted to be an MSP using machine-learning techniques \citep{SazParkinson+2016}; although, a pulsar search using 8-min GBT observations at 2\,GHz did not detect pulsations \citep{Sanpa-arsa16}.
This adds to the demonstration of the high quality of the TULIPP candidates identified and the potential for follow-up observations at higher frequencies. 

%\textbf{Pulsar radio beam geometry:} 
Alternatively, there is a possibility that we cannot detect pulsations from sources because the radio emission does not vary with rotational spin, including neutron stars/pulsars 
%\LEt{reminder notes 1\&2}
with aligned rotation and magnetic axes pointed directly towards our line of sight. 
In this case, searching the polarisation data using the Stokes $Q$, $U,$ and $V$ parameters may provide an additional avenue, as these may be more likely to vary with spin phase in this case \citep[e.g. following the rotating vector model, RVM;][]{rc69}. 
However, this geometry may also reduce the average degree of linear polarisation that we measure using the LoTSS images, and it may not apply to significantly polarised sources \citep[e.g.][]{Crawford+2000}.

It is unlikely that the polarised LoTSS sources were not detected, because the pulsar radio emission beam does not intersect our line of sight. However, it is still debated whether there is unpulsed continuum emission from pulsars \citep[e.g.][for PSR\,J0218+4232]{Navarro+1995}, and unpulsed emission detected may be due to pulsar wind nebulae (plerions), supernova remnants, or other structures in the ISM \citep[e.g.][and references therein]{Ravi+2018}.

%\textbf{Other radio sources:}
High fractional linear and/or circular polarisation is a well-known continuum property of pulsars and has resulted in previous pulsar discoveries as outlined in \S\,\ref{sec:introduction}, and it is mostly unique for radio sources. However, other sources of polarised emission exist, including rare flare stars, star-planet interactions, ultracool dwarfs, radio galaxies, and blazars \citep[e.g.][]{O'Sullivan+2013,Vedantham+2020,Pritchard+2021,Callingham+2021}. These linear and circularly polarised sources will be further discussed in O'Sullivan et al., in prep. and Callingham et al., in prep., respectively. 
For the linearly polarised candidates that we discovered/detected as pulsars, both have measured RMs $<$65\% of that expected from the Galactic line-of-sight, Table~\ref{tab:sources}. 
Although five linearly polarised candidates have measured RMs larger than that expected from the Galactic line of sight, only ILT\,J042737.00+684536.9 is over 1$\upsigma$ larger, taking into account the uncertainties.  This source may be, e.g, a compact radio galaxy, as it also has a relatively low linear polarisation fraction. 
However, RMs towards pulsars show varying agreement with the RMs from the entire Galactic line of sight \citep[e.g.][]{sbg+19}. 
Notably, a candidate with the highest fractional linear polarisation and two candidates with the highest circular polarisation fractions were not detected as pulsars, making them particularly interesting for further investigation into the source(s) of their emission.

\subsection{Future prospects }
 
The TULIPP pulsar discovery rate is two pulsars, one of which is an MSP, and the pulsar detection rate is a further five pulsars (including 3 MSPs), in 21\,h of LOFAR observations. 
However, over 40 known pulsars, half of which are MSPs, were identified as polarised point sources in the LoTSS data and excluded from the TULIPP survey observations.
The LOTAAS discovery rate is 74 pulsars (including 2 MSPs), and the pulsar detection rate is a further 311 pulsars (including 6 MSPs), in 1910\,h of observations \citep{Sanidas+2019}. 
In terms of pulsars discovered per hour of LOFAR observations, the first year of the TULIPP survey has been more efficient, particularly for short-period MSPs, compared to the LOTAAS all-sky survey. 
LOTAAS has surveyed the northern sky above a $0\degr$ declination, while TULIPP has observed 30 $\sim3\farcm5$ fields containing polarised sources that were catalogued using $\sim150$ LoTSS fields that cover $\sim2400$\,deg$^{2}$ of sky.
As the TULIPP survey progresses, we will be able to more fully compare and contrast these complementary surveys, in terms of efficiency in telescope time and sky area surveyed. 

The LOFAR discovery of the new pulsars, PSRs\,J1049+5822 and J1602+3901, by following up linearly and circularly polarised point sources identified in LOFAR images adds to the pulsar discovered in the 3C\,196 observations taken by the Epoch of Reionisation LOFAR Key Science Project \citep{Jelic+2015}. 
These and similar discoveries using other telescopes summarised in \S\,\ref{sec:introduction} show the value of targeted follow-up of polarised point sources in continuum imaging as a method of discovering new pulsars. 
However, this method of following up linearly and circularly polarised sources in the imaging domain could miss pulsars with pulse profiles that show large swings in the P.A., show swings between handedness in circular polarisation, as expected from the core of the pulsar emission beam \citep[][and references therein]{Everett+2001,Melrose+2004}, or that are not significantly polarised at all.  
Furthermore, as the LoTSS survey expands towards lower Galactic latitudes, we may see a decrease in the number density of linearly polarised sources, due to larger |RMs|, although the circularly polarised emission will not be affected.
We can investigate if this effects detecting known pulsars, and, therefore, the prospects for identifying pulsar candidates. 
Furthermore, there are alternative methods and prospects for identifying promising pulsar candidates in continuum images including following up sources that show variability, steep-spectral indices, or associations with ISM structures such as supernovae or pulsar wind nebulae \citep[e.g.][]{Dai+2017,Dai+2018,Schinzel+2019,Hyman+2021}.
 
PSR\,J1049+5822 appears to be a reasonably typical non-recycled pulsar, located in the middle of the pulsar island in the $P-\dot{P}$ diagram. 
Thus, it seems there are more pulsars to be discovered in this region of the pulsar parameter space, as expected when using increasingly sensitive observations from SKA precursors/pathfinders.
This will continue to help us understand the Galactic population of pulsars, and the potential discoveries in future pulsar surveys using the SKA \citep[e.g.][]{Keane+2015,Xue+2017}.

All of the LoTSS DR2 data have now been processed through the \textsc{RM Grid} pipeline and will be summarised in O'Sullivan et al., in prep. 
TULIPP is now a long-term LOFAR observing project, and we have more recently observed tens of more linearly polarised and circularly polarised pulsar candidate sources. 
As the LoTSS observations continue to observe larger sky areas, particularly extending further towards the Galactic plane, and eventually covering the entire northern sky \citep{sth+19}, we can expect additional promising pulsar candidates to follow up using LOFAR beam-formed observations. 
Following this, the upgrade to LOFAR2.0 will enable more sensitive sky surveys by facilitating observations using all low band antenna dipoles and HBAs simultaneously, and commensal imaging and beam-formed observations.  
There are additional prospects to identifying candidates, including using lower flux-density-limit thresholds for the polarised LoTSS sources, and increasing the sensitivity of the polarisation images by processing the LoTSS data products at a $6\arcsec$ resolution and creating mosaicked images from the individual fields.

More generally, we have demonstrated that a targeted pulsar search of polarised point sources at low frequencies can successfully and relatively efficiently discover pulsars, including MSPs.
Therefore, this method shows promise for other SKA precursor/pathfinder facilities, and, in future, the SKA. 
Indeed, such a targeted pulsar search accompanying a sensitive SKA polarisation imaging survey may be particularly productive in discovering MSPs, while an all-sky pulsar search using coherent de-dispersion remains computationally infeasible. 
In the meantime, understanding pulsar properties in continuum images can provide valuable input for even more efficient classification of radio sources and the identification of promising pulsar candidates, for example, as a potential classifier in machine learning approaches \citep{Tan+2018,Galvin+2020,Mostert+2021}.

\section{Summary and conclusions}\label{sec:conclusions}

We describe the TULIPP (targeted search using LoTSS images for polarised pulsars) pulsar survey and present the results from the first year of LOFAR observations in observing semesters LC12 and LC13 (2019 June 1 -- 2020 May 31). We summarise this work as follows.

\begin{itemize}
 \item We identified 30 promising pulsar candidates (excluding over 40 known pulsars) using LOFAR Two-Metre Sky Survey (LoTSS) linear and circular polarisation images. The selection criteria included the requirements that sources were unresolved in the 20 and 6\arcsec\ total intensity images, with greater than 7\% linear or 1\% circular fractional polarisation. 
 \item We followed up nine linearly polarised sources and 21 circularly polarised pulsar-candidate sources using 21\,h of LOFAR tied-array beam-formed observations centred at 149\,MHz. The data were processed using three independent pulsar search pipelines.   
\item During the first year of the TULIPP survey, we discovered two pulsars: PSR\,J1049+5822 is a `slow' pulsar with spin period of $P=0.7276$\,s and dispersion measure of $\mathrm{DM}=12.364$\,pc\,cm$^{-3}$. PSR\,J1602+3901 is an MSP with spin period of $P=0.0037$\,s and dispersion measure of $\mathrm{DM}=17.259$\,pc\,cm$^{-3}$.
\item We also independently re-discovered (detected) a further five known pulsars that did not have well-localised sky positions: `slow' pulsars  
PSRs\,J0220+3626, and J2227+3038; and MSP PSRs\,J0742+4110, J1630$+$3550 and J2212+2450.
\item We conducted further observations for PSRs\,J1049+5822 and J1630$+$3550 and presented the parameters measured using pulsar timing. We found that PSR\,J1630$+$3550 is part of a 7.58-h orbital period binary system with a low-mass companion, suggesting it is a `black-widow' binary MSP system.
\item We presented the polarisation pulse profiles of the pulsars we detected in the TULIPP data and further discussed their properties. We also discussed potential reasons for non-detections of pulsars towards the other candidate sources, the future prospects for the TULIPP survey, and using this methodology for future targeted pulsar searches including those using SKA observations.
   \end{itemize}
   
Our results demonstrate that pulsars, and particularly MSPs, can be relatively efficiently discovered using this targeted pulsar search method. 

%%%%%%%%%%%%%%%%%

\begin{acknowledgements}
  This paper is based on data obtained with the International LOFAR
  Telescope (ILT) under project codes LC12\_003, LC13\_006 as well as LT10\_015 and LT14\_005. 
  LOFAR \citep{hwg+13} is the Low Frequency
  Array designed and constructed by ASTRON. It has observing, data
  processing, and data storage facilities in several countries,
  that are owned by various parties (each with their own funding
  sources), and that are collectively operated by the ILT
  foundation under a joint scientific policy. The ILT resources
  have benefitted from the following recent major funding sources:
  CNRS-INSU, Observatoire de Paris and Universit\'{e}
  d'Orl\'{e}ans, France; BMBF, MIWF-NRW, MPG, Germany; Science
  Foundation Ireland (SFI), Department of Business, Enterprise and
  Innovation (DBEI), Ireland; NWO, The Netherlands; The Science and
  Technology Facilities Council, UK.  
  We thank the anonymous referee and Phil Edwards for comments that improved the manuscript.
  We thank Dale Frail and Paul Ray for providing information on gamma-ray and Fermi Pulsar Search Consortium data.
  J.W.T.H., C.G.B. and V.I.K. acknowledge funding from an NWO Vidi fellowship and ERC Starting Grant ``DRAGNET'' (337062).  J.W.T.H. also acknowledges funding from an NWO Vici grant ("AstroFlash"; VI.C.192.045).
  M.H. acknowledges funding from the European Research Council (ERC) under the European Union's Horizon 2020 research and innovation programme (grant agreement No 772663).
  This work used the \textsc{Python} packages: \textsc{Astropy} \citep{astropy:2013,astropy:2018}, \textsc{Matplotlib} \citep{Hunter:2007}, \textsc{NumPy} \citep{harris2020array}.  
  This research has made use of NASA's Astrophysics Data System Bibliographic Services.
\end{acknowledgements}

\bibliographystyle{aa} 
\bibliography{references} 

\end{document}